\documentclass[preprint,journal]{vgtc}       

\ifpdf
  \pdfoutput=1\relax                   
  \usepackage{graphicx}                
  \DeclareGraphicsExtensions{.pdf,.png,.jpg,.jpeg} 
\else
  \usepackage{graphicx}                
  \DeclareGraphicsExtensions{.eps}     
\fi%

\graphicspath{{figures/}}

\usepackage{microtype}                 
\PassOptionsToPackage{warn}{textcomp}  
\usepackage{textcomp}                  
\usepackage{mathptmx}                  
\usepackage{times}                     
\usepackage{cite}                      
\usepackage{tabu}                      
\usepackage{booktabs}                  

\usepackage{subfig}
\usepackage{amsmath}
\usepackage{amssymb}
\usepackage{siunitx}
\usepackage{bm}
\usepackage{booktabs}

\newcommand{\SetFont}[1]{\mathbb{#1}}
\newcommand{\VectorFont}[1]{#1}
\newcommand{\N}{\SetFont{N}}
\newcommand{\NZero}{\SetFont{N}_0}
\newcommand{\R}{\SetFont{R}}
\DeclareMathOperator{\erf}{erf}
\newcommand{\Transpose}[1]{#1^\mathsf{T}}
\newcommand{\SampleCount}{n}
\newcommand{\SampleIndex}{i}
\newcommand{\DataVectorCount}{m_v}
\newcommand{\DataVectorIndex}{j}
\newcommand{\DataScalarCount}{m_s}
\newcommand{\DataScalarIndex}{j}
\newcommand{\DataAnyIndex}{j}
\newcommand{\OtherDataAnyIndex}{k}
\newcommand{\DataAnyCount}{m_u}
\newcommand{\DataPosition}[1]{\VectorFont{v}_{#1,0}}
\newcommand{\DataVector}[2]{\VectorFont{v}_{#1,#2}}
\newcommand{\DataScalar}[2]{s_{#1,#2}}
\newcommand{\DataComplete}[1]{\VectorFont{u}_{#1}}
\newcommand{\DataAny}[2]{\DataComplete{#1,#2}}
\newcommand{\FrameIndex}{t}
\newcommand{\ClusterIndexSet}{\SetFont{I}}
\newcommand{\ClusterIndexSubSet}{\SetFont{I_S}}
\newcommand{\DimensionIndexSet}{\SetFont{J}}
\newcommand{\GMMIndex}{g}
\newcommand{\GaussianCount}[1]{n_{#1}}
\newcommand{\GaussianIndex}{l}
\newcommand{\GaussianWeight}[2]{w_{#1,#2}}
\newcommand{\GaussianMean}[2]{\VectorFont{\mu}_{#1,#2}}
\newcommand{\GaussianCovariance}[2]{\VectorFont{\Sigma}_{#1,#2}}
\newcommand{\GaussianDensity}[1]{\rho_{#1}}
\newcommand{\AttributePosition}{\VectorFont{p}}
\newcommand{\LogLikelihood}{L_p}
\newcommand{\GMMFreeParams}{k_{\text{GMM}}}

\newcommand{\SingleGaussianWeight}{\VectorFont{w}}
\newcommand{\SingleGaussianMean}{\VectorFont{\mu}}
\newcommand{\SingleGaussianCovariance}{\VectorFont{\Sigma}}
\newcommand{\RayOrigin}{\VectorFont{o}}
\newcommand{\RayDirection}{\VectorFont{d}}
\newcommand{\RayVariable}{x}
\newcommand{\TransformedRayVariable}{y}
\newcommand{\IntervalBegin}{a}
\newcommand{\IntervalEnd}{b}
\newcommand{\GaussianIntegral}[2]{I(#1,#2)}
\newcommand{\IntegralAuxiliary}{c}
\newcommand{\IntegralAuxiliaryOffsetDirection}{c_{\RayOrigin,\RayDirection}}
\newcommand{\IntegralAuxiliaryDirection}{c_{\RayDirection,\RayDirection}}
\newcommand{\GaussianOneDimMean}[2]{\VectorFont{\mu}_{#1,#2}}
\newcommand{\GaussianOneDimVariance}[2]{\VectorFont{\sigma^2}_{#1,#2}}

\newcommand{\TF}{f}
\newcommand{\PCPVariableX}{x}

\newcommand{\PCPInterpolatedMean}[1]{\mu(\PCPVariableX)}
\newcommand{\PCPInterpolatedVariance}[1]{\sigma^2(\PCPVariableX)}
\newcommand{\Density}{\rho}
\newcommand{\DensityScale}{\lambda}
\newcommand{\ClusterIndexSetCurrent}[1]{\ClusterIndexSet_{\FrameIndex,{#1}}}
\newcommand{\ClusterIndexSetNext}[1]{\ClusterIndexSet_{\FrameIndex+1,{#1}}}
\newcommand{\ClusterCountCurrent}{n_\FrameIndex}
\newcommand{\ClusterCountNext}{n_{\FrameIndex+1}}
\newcommand{\ClusterDegreeOfInterestCurrent}[1]{d_{\FrameIndex,{#1}}}
\newcommand{\ClusterDegreeOfInterestNext}[1]{d_{\FrameIndex+1,{#1}}}
\newcommand{\ClusterIndex}{k}
\newcommand{\OtherClusterIndex}{l}

\newcommand{\Wasserstein}{W}
\newcommand{\CdfDataSamples}{F_{\ClusterIndexSet{}}}
\newcommand{\CdfGMM}{F_{\GMMIndex}}

\newcommand{\OutlierPercentage}{p_o}

\ifdefined\ShowRevision
	\usepackage[normalem]{ulem}
	
	\newcommand{\revi}[1] {{\color{blue}#1}}
	\newcommand{\revirm}[1] {\revi{\sout{#1}}}
\else
	\newcommand{\revi}[1] {#1}
	\newcommand{\revirm}[1] {}
\fi

\ieeedoi{10.1109/TVCG.2020.3030379}

\onlineid{1115}

\vgtccategory{Research}
\vgtcpapertype{System}

\title{Visual Analysis of Large Multivariate Scattered Data using Clustering and Probabilistic Summaries}

\author{Tobias~Rapp, Christoph~Peters, and~Carsten~Dachsbacher}
\authorfooter{
\item
Tobias Rapp, Christoph Peters, Carsten Dachsbacher are with Karlsruhe Institute of Technology. E-mail: {tobias.rapp, christoph.peters, dachsbacher}@kit.edu.

}

\abstract{
	Rapidly growing data sizes of scientific simulations pose significant challenges for interactive visualization and analysis techniques.
	In this work, we propose a compact probabilistic representation to interactively visualize large scattered datasets.
	In contrast to previous approaches that represent blocks of volumetric data using probability distributions, we model clusters of arbitrarily structured multivariate data.
	In detail, we discuss how to efficiently represent and store a high-dimensional distribution for each cluster. We observe that it suffices to consider low-dimensional marginal distributions for two or three data dimensions at a time to employ common visual analysis techniques. Based on this observation, we represent high-dimensional distributions by combinations of low-dimensional Gaussian mixture models.
	We discuss the application of common interactive visual analysis techniques to this representation. In particular, we investigate several frequency-based views, such as density plots in 1D and 2D, density-based parallel coordinates, and a time histogram. We visualize the uncertainty introduced by the representation, discuss a level-of-detail mechanism, and explicitly visualize outliers. Furthermore, we propose a spatial visualization by splatting anisotropic 3D Gaussians for which we derive a closed-form solution.
	Lastly, we describe the application of brushing and linking to this clustered representation. Our evaluation on several large, real-world datasets demonstrates the scaling of our approach.
} 


\teaser{
	\centering
	\includegraphics[width=\linewidth]{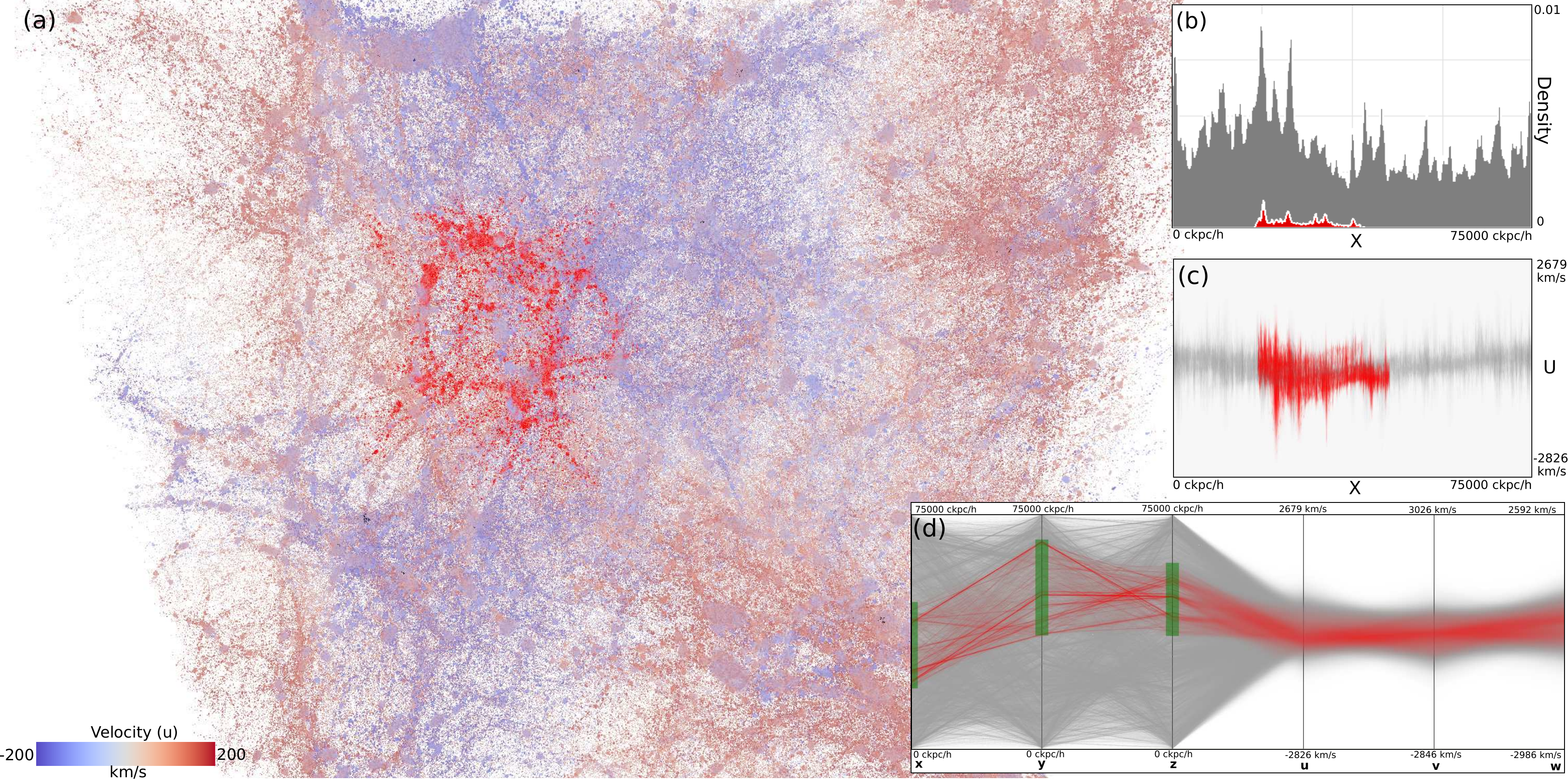}
	  \caption{
		Our probabilistic summary of a cosmological dataset represents 2.6 billion particles partitioned into 5.3~million clusters. We model each cluster using combinations of low-dimensional Gaussian mixture models. This allows us to interactively visualize the position of particles by splatting 3D Gaussians (a) and to create density-based 1D and 2D plots, depicted in (b) and (c). A density-based parallel coordinate plot is shown in (d). All of those views support interactive navigation and exploration by brushing (red) and linking. We render this massive dataset in 28 ms on an NVIDIA GTX 1080 Ti at a resolution of $1920\times1080$.}
	\label{fig:teaser}
}

\vgtcinsertpkg

\begin{document}


\firstsection{Introduction}

\maketitle

The field of scientific visualization is confronted with rapidly growing amounts of data, including multivariate and time-dependent data. Interactive visual analysis~\cite{Weber2014} approaches have been established as a powerful approach to facilitate knowledge discovery in complex datasets.
However, growing data sizes make the interactive exploration increasingly difficult or even impossible for some datasets.

To deal with large amounts of data, recent approaches employ probabilistic data summaries~\cite{Wang2017, Dutta2017, Dutta2017a, Hazarika2019} to represent blocks of data as probability distributions.
These approaches have been mostly limited to univariate, volumetric data. In this work, we propose a representation that supports arbitrarily structured, time-dependent, and multivariate data defined in a two- or three-dimensional spatial domain.
To this end, the data needs to be partitioned, i.e.\ clustered into spatially coherent regions. In each cluster, we make use of Gaussian mixture models (GMMs) to compactly represent a probability distribution of the data using a weighted combination of Gaussian components. However, multivariate data requires modeling high-dimensional distributions, which suffer from the curse of dimensionality.
Our approach is based on the observation that representations of low-dimensional marginal distributions suffice to analyze and visualize the data. All common visualizations, such as scatter plots, histograms, and parallel coordinate plots, require only 1D or 2D distributions. The exception is the spatial domain of scattered data in 3D for which we employ a 3D distribution. Thus, we model the marginal distributions of all individual dimensions and pairs of dimensions as well as the spatial 3D distribution.

For large data, common item-based visualizations, such as scatter and parallel coordinate plots, are challenged by overdraw and cluttering. Frequency-based visualizations are a viable alternative in this case~\cite{Novotny2006, Lampe2011}. Density estimation~\cite{Silverman2018} is a frequency-based approach commonly used in statistics. However, its usage in interactive visualization has been limited due to performance considerations.
Although our approach supports all common visualization techniques, it is especially well suited to density-based techniques since our modeled distributions are already an estimate of density. We discuss the efficient visualization and interaction with density-based plots using our compact representation.
Additionally, we consider time-dependent histograms that would otherwise be infeasible to produce for large datasets. In this view, we can interactively brush over different time steps.
To visualize the uncertainty introduced by our data representation, we propose an error metric based on the cumulative distribution function, similar to statistical goodness of fit tests. A level-of-detail mechanism allows scientists to drill down on interesting or uncertain regions in the data. Additionally, we discuss the explicit visualization of outliers, which are not handled well by density-based visualizations.

Our last contribution is the visualization of spatial density distributions. Since drawing and rendering samples from the GMMs would be infeasible for large, scattered datasets, we directly render 3D Gaussians. We derive a closed-form solution to integrate anisotropic Gaussians using a splatting approach. Back-to-front splatting has the disadvantage that it assumes non-overlapping Gaussians. Therefore, we employ moment-based order-independent transparency~\cite{Muenstermann2018} for datasets where this is not an acceptable assumption.

\noindent
To summarize, our main contributions are:
\begin{itemize}
	\item We define compact data representations based on probabilistic models of low-dimensional marginal distributions for scattered, multivariate data,
	\item We describe interactive visual analysis techniques based on our probabilistic data summaries,
	\item We efficiently visualize scattered, overlapping, anisotropic 3D Gaussians.
\end{itemize}

\section{Related Work}
\label{sec:Related}

In this section, we discuss previous work on probabilistic data modeling, density-based visualizations, and the rendering of 3D Gaussians.

\subsection{Probabilistic Modeling of Large Data}

Several probabilistic approaches to represent large volumetric, univariate datasets have been proposed. Thompson et al.~\cite{Thompson2011} describe \emph{hixels}, a data representation that stores a histogram per block of voxels. \revi{Liu et al.~\cite{Liu2012} discuss volume rendering using per voxel Gaussian mixture models.}
Sicat et al.~\cite{Sicat2014} construct a multi-resolution volume from sparse probability density functions defined in the 4D domain comprised of the spatial and data range. To visualize and analyze large volumetric data, Wang et al.~\cite{Wang2017} employ a spatial GMM in addition to a value distribution in each data block.
For in-situ feature analysis of time-varying data, Dutta et al.~\cite{Dutta2017a} perform incremental GMM estimation instead of expectation maximization, which is traditionally used to estimate the parameters of a mixture model.
By design, none of these approaches is applicable to more than four dimensions.
Dutta et al.~\cite{Dutta2017} model a single Gaussian or a GMM with a fixed number of components to each univariate value distribution in a cluster of the data. The authors compare several clustering techniques to determine homogeneous regions in volumetric data.
Since an optimal clustering of the data is generally domain or application specific, we do not make any assumptions about the clustering procedure. \revi{Our method overcomes the limitation to low-dimensional data by working with low-dimensional GMMs for all relevant combinations of dimensions. We also introduce a fast and adaptive selection of the number of GMM components.}

For parameter studies in cosmological simulations, Wang et al.~\cite{Wang2019} store GMMs as a prior knowledge to reconstruct high-resolution datasets from multiple prior simulation runs. \revi{Li et al.~\cite{Li2020} reduce cosmological simulation data in-situ by subdividing space using a k-d tree and estimating particle density using a GMM in each leaf node. During the analysis stage, particles are sampled from the GMMs.}
Hazarika et al.~\cite{Hazarika2018} propose a copula-based uncertainty modeling approach to represent a multivariate distribution using different types of univariate distributions, including GMMs, separately from their interrelation. To summarize large-scale multivariate volumetric data, a copula-based analysis framework has been introduced~\cite{Hazarika2019}.
This approach is the first to address the modeling of multivariate data, but the Gaussian copula function limits the correlations between dimensions to a single Gaussian.
\revi{Whilst we similarly decompose a high-dimensional model into more manageable low-dimensional models, we do not share this limitation.}
Moreover, the approaches of Hazarika et al.\ and Li et al.\ require\revirm{s} sampling, which hinders \revi{the} application to interactive visual analysis, especially for rendering scattered data.
\revi{Similarly, we do not perform subsampling of scattered data~\cite{Woodring2011, Rapp2019} for data reduction since our GMMs already estimate density, which we use directly in our density-based visualizations.}

\subsection{Density-Based Scatter and Parallel Coordinate Plots}

Scatter and parallel coordinate plots can be used to visualize multivariate data.
For large data, these item-based visualizations are challenged by overdraw and visual clutter. Instead of drawing discrete glyphs, density estimation methods reconstruct and visualize a continuous density of data values.
For scatterplots, a simple form of density estimation is to draw individual points semi-transparently using alpha blending. Histograms and hexagonal binning are often employed to convey frequency information, but can lead to aliasing due to their discrete nature. The concept of histograms has also been extended to parallel coordinate space~\cite{Artero2004, Novotny2006, Blaas2008}.
Although kernel density estimation would allow for an improved reconstruction of continuous density, it is computationally expensive.
Splatterplots~\cite{Mayorga2013} perform kernel density estimation to avoid overdraw, but explicitly add representative outliers. We estimate density using GMMs and similarly support the explicit visualization of outliers.

In the field of scientific visualization, continuous scatter and parallel coordinate plots have been introduced~\cite{Bachthaler2008, Heinrich2009} to construct density plots by considering the topology and interpolation of data samples in their spatial domain.
Despite optimizations~\cite{Heinrich2011}, this remains a computationally challenging approach that is unsuited for the interactive analysis of large-scale datasets.
\revirm{
To transform bivariate Gaussians to parallel coordinate space, Miller and Wegman~[46] present a closed-form solution. We make use of this formulation to efficiently create density-based parallel coordinate plots from Gaussian distributions.
Lastly, for brushing uncertain data, we extend the approach of Feng et al.~[9] to our data representation.}

\subsection{Rendering 3D Gaussians}
The encoding of scattered, unstructured, or large volumetric data using radial basis functions (RBF) has been an active research topic~\cite{Co2003, Jang2004, Jang2006, Weiler2005}. This involves the rendering of isotropic and anisotropic Gaussian kernels~\cite{Neophytou2006, Juba2007, Hong2009}.
In detail, Zwicker et al.~\cite{Zwicker2001} discuss splatting of elliptical Gaussians by approximating the footprint after perspective projection. They extend their splatting approach by combining the reconstruction with a low-pass kernel, which could be similarly applied to our approach.
In contrast to previous work, we derive a closed-form solution to integrate a Gaussian kernel along a ray. This enables us to efficiently splat large amounts of Gaussians without requiring expensive precomputation.
Note that we consider three-dimensional Gaussians defined by a mean and covariance matrix, which makes the use of a view-independent look-up table infeasible.
Additionally, we employ moment-based order-independent transparency~\cite{Muenstermann2018} to address the short-comings of back-to-front splatting.
\revi{Our method could also be employed during volume ray casting~\cite{Knoll2014}.}

\section{Probabilistic Summaries}

\begin{figure}
	\centering
	\subfloat[Clustered data]{%
		\includegraphics[width=0.49\linewidth]{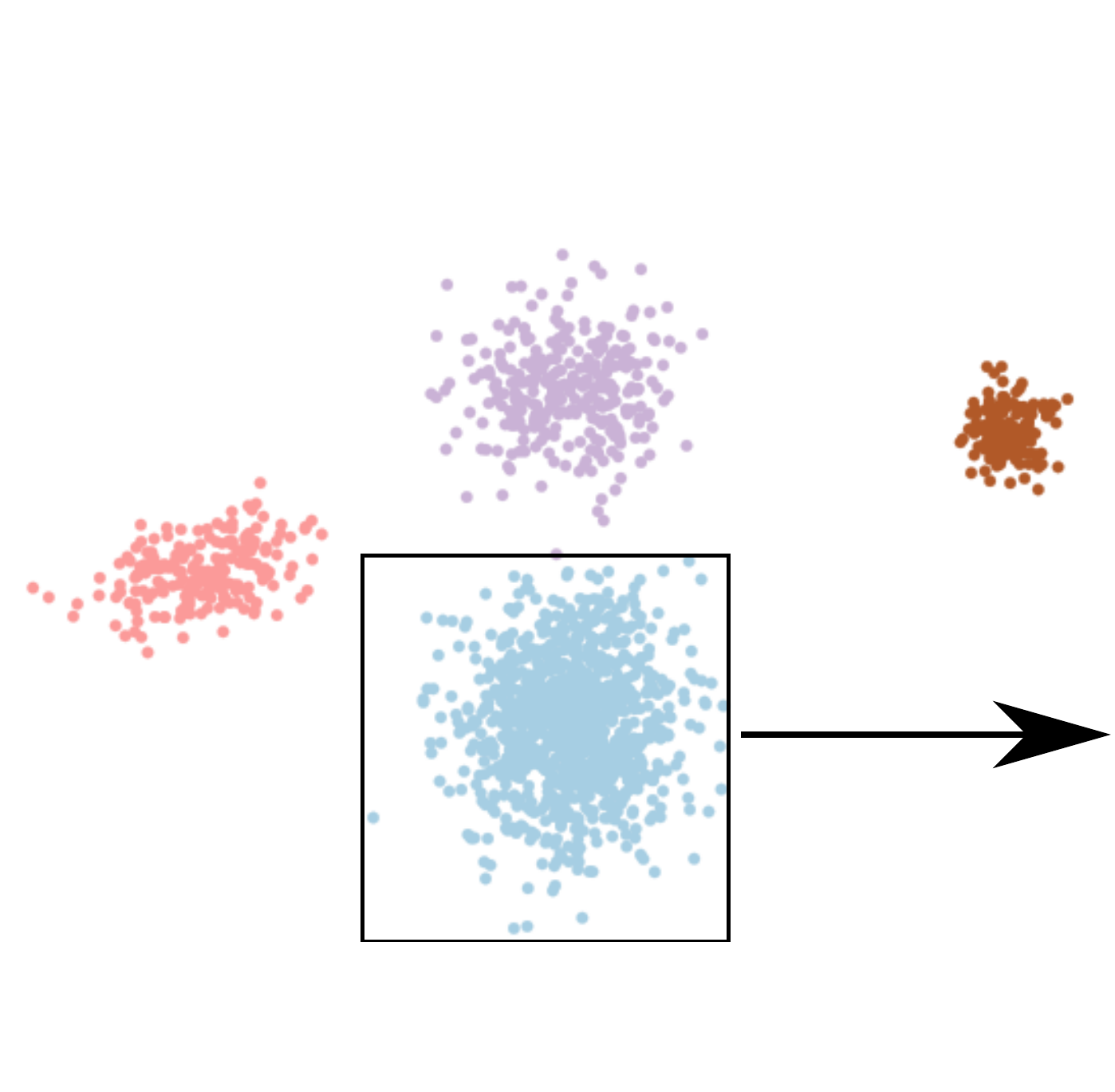}
	}
	\hfill
	\subfloat[Scatter plot matrix]{%
		\includegraphics[width=0.49\linewidth]{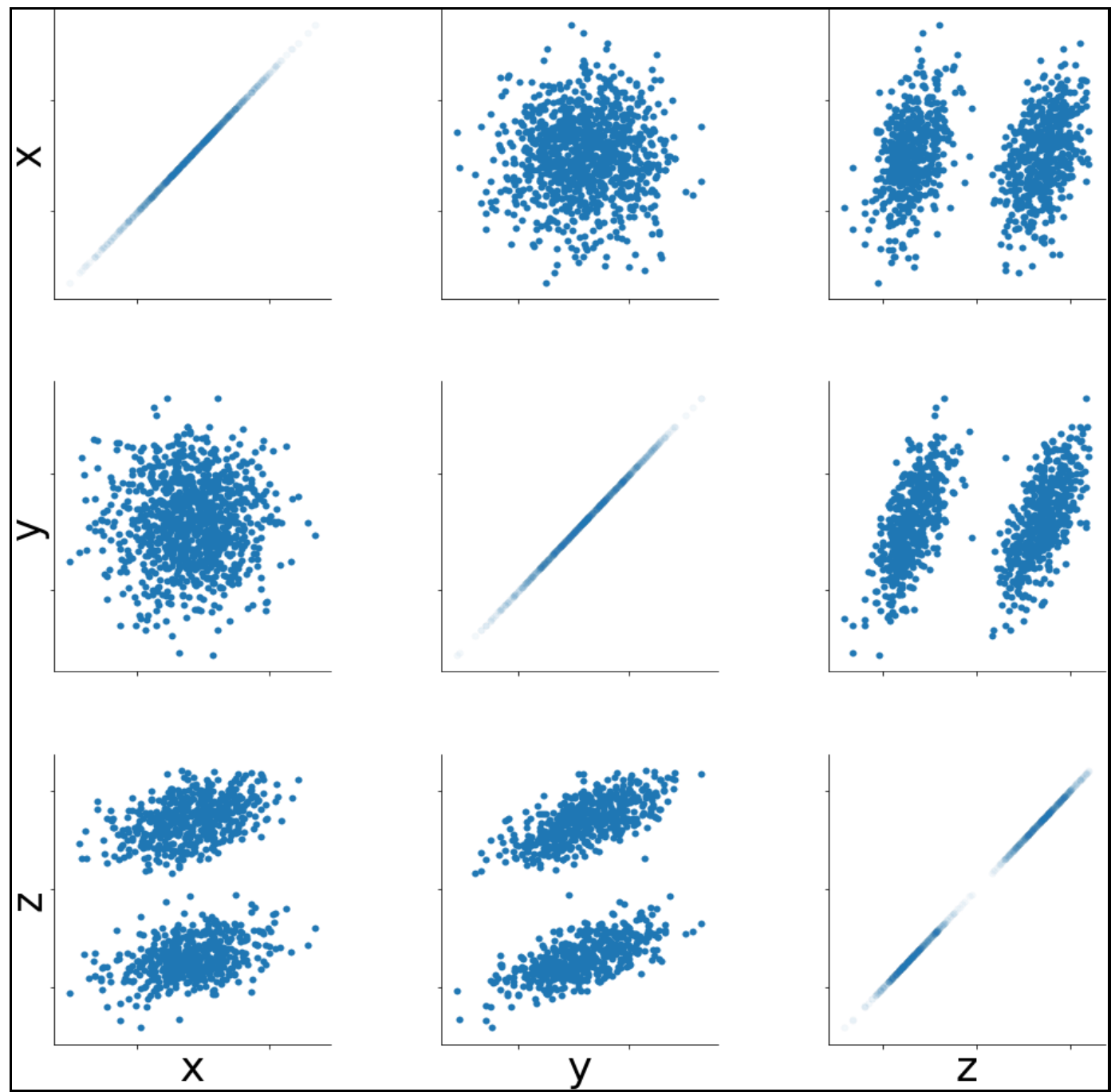}
	}
	\vspace{-1mm}
	\caption{From a given clustering of the data (a), we model each cluster using combinations of low-dimensional distributions, similar to a scatter plot matrix (b). }
	\label{fig:ClusterOverview}
	\vspace{-4mm}
\end{figure}

In this work, we describe the creation of probabilistic data summaries for multivariate, scattered data. We assume that the data is clustered into spatially coherent regions~\cite{Dutta2017}. In \autoref{sec:Results}, we discuss both domain specific and standard clustering techniques for scattered data.
Similar to previous work, we employ Gaussian mixture models to represent data distributions in each cluster. However, these have not been applied to multivariate data. High-dimensional Gaussian mixture models require immense computational effort and due to the curse of dimensionality, there are not enough samples to cover a multi-dimensional space extensively.

\revi{Our approach is based on the observation that we do not require} more than three data-dimensions at once \revi{to employ common interactive visual analysis techniques}. In fact, the visualization of the spatial distribution is the only aspect considering correlations of three dimensions. Therefore, our approach is to only generate GMMs for the relevant combinations of dimensions. By default, these are all individual dimensions, all pairs of dimensions (cf. \autoref{fig:ClusterOverview}) and all vectorial attributes. As for high-dimensional GMMs, the storage cost grows quadratically with the number of dimensions. 
To better reason about our approach, we first introduce it more formally.

\subsection{Data Model}

Our data consists of $\SampleCount\in\N$ samples. Each sample is associated with a position in 3D space, $\DataVectorCount-1\in\NZero$ additional vectorial attributes and $\DataScalarCount\in\NZero$ scalar attributes. Of course, the approach is also applicable to scattered data in 1D or 2D space. We denote the data for sample $\SampleIndex\in\{0,\ldots,\SampleCount-1\}$ by:
\begin{itemize}
	\item $\DataPosition{\SampleIndex}\in\R^{1\times3}$ for the position,
	\item $\DataVector{\SampleIndex}{\DataVectorIndex}\in\R^{1\times3}$ for vectorial attribute $\DataVectorIndex\in\{1,\ldots,\DataVectorCount-1\}$,
	\item $\DataScalar{\SampleIndex}{\DataScalarIndex}\in\R$ for scalar attribute $\DataScalarIndex\in\{0,\ldots,\DataScalarCount-1\}$.
\end{itemize}
To define our probabilistic summaries, we concatenate all attributes for sample $\SampleIndex\in\{0,\ldots,\SampleCount-1\}$ into a single vector with $\DataAnyCount:=3\DataVectorCount+\DataScalarCount$ entries to enable linear indexing:
$$\DataComplete{\SampleIndex}:=(\DataPosition{\SampleIndex},\DataVector{\SampleIndex}{1},\ldots,\DataVector{\SampleIndex}{\DataVectorCount-1},\DataScalar{\SampleIndex}{0},\ldots,\DataScalar{\SampleIndex}{\DataScalarCount-1})\in\R^{1\times\DataAnyCount}.$$
GMMs are generated for each given cluster $\ClusterIndexSet\subseteq\{0,\ldots,\SampleCount-1\}$ and for each relevant combination of dimensions. First, we generate 1D models for each attribute $\DataAnyIndex\in\{0,\ldots,\DataAnyCount-1\}$:
$$(\DataAny{\SampleIndex}{\DataAnyIndex})_{\SampleIndex\in\ClusterIndexSet}\in\R^{|\ClusterIndexSet|\times1}.$$
Then, we generate 2D models for each pair of dimensions \linebreak{}$\DataAnyIndex,\OtherDataAnyIndex\in\{0,\ldots,\DataAnyCount-1\}$ with $\DataAnyIndex<\OtherDataAnyIndex$:
$$(\DataAny{\SampleIndex}{\DataAnyIndex},\DataAny{\SampleIndex}{\OtherDataAnyIndex})_{\SampleIndex\in\ClusterIndexSet}\in\R^{|\ClusterIndexSet|\times2}.$$
Finally, we generate 3D models for each vectorial attribute \linebreak{}$\DataVectorIndex\in\{0,\ldots,\DataVectorCount-1\}$:
$$(\DataVector{\SampleIndex}{\DataVectorIndex})_{\SampleIndex\in\ClusterIndexSet}\in\R^{|\ClusterIndexSet|\times3}.$$
%
%
Of course, the generation of GMMs for particular combinations of attributes may be skipped if the analysis of their mutual dependence is of no interest in a specific application.

Our probabilistic summary is the combination of all these low-dimensional GMMs for all clusters.
They capture all information needed for common interactive visual analysis techniques \revi{but limit the analysis of higher dimensional correlations}. By modeling only low-dimensional distributions, the curse of dimensionality does not apply.
\revi{Ultimately, this limitation enables us to create reliable models of multivariate data.}

\subsection{Gaussian Mixture Models}
\label{sec:Model:GMMs}

\begin{figure}
	\centering
	\includegraphics[width=\linewidth]{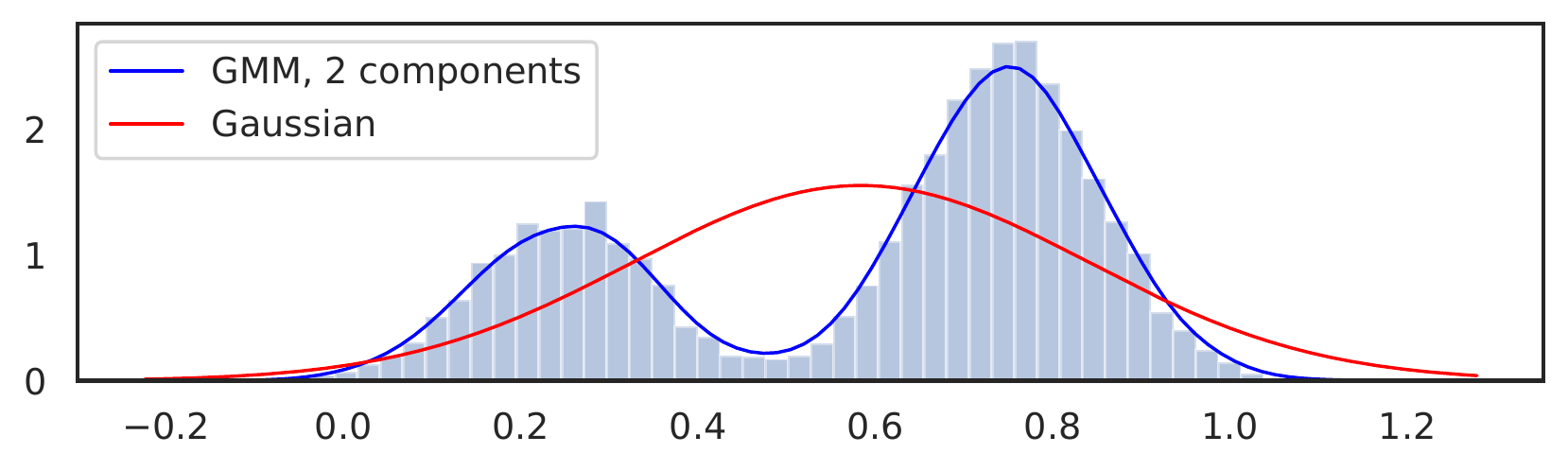}
	\vspace{-6mm}
	\caption{To the distribution shown by the histogram, we have fitted a Gaussian (red) and a mixture of two Gaussians (blue). In this example, the GMM better models the data.}
	\label{fig:GaussiansGMM}
	\vspace{-3mm}
\end{figure}

We use GMMs because they offer a compact and efficient representation of the target distributions and have been employed successfully for \revi{modeling low-dimensional distributions} in previous work~\cite{Dutta2017, Dutta2017a, Wang2017}. In the following, we provide more details on our fitting procedure.

We generate a GMM for each combination of a cluster $\ClusterIndexSet$ and a relevant subset of attributes $\DimensionIndexSet\subseteq\{0,\ldots,\DataAnyCount-1\}$. As explained above, the number of attributes $|\DimensionIndexSet|$ is one, two, or three. A GMM is indexed by a pair $\GMMIndex:=(\ClusterIndexSet,\DimensionIndexSet)$ and consists of $\GaussianCount{\GMMIndex}\in\N$ weighted Gaussians. Gaussian $\GaussianIndex\in\{0,\ldots,\GaussianCount{\GMMIndex}-1\}$ is given by its weight $\GaussianWeight{\GMMIndex}{\GaussianIndex}>0$, its mean $\GaussianMean{\GMMIndex}{\GaussianIndex}\in\R^{|\DimensionIndexSet|}$ and its covariance $\GaussianCovariance{\GMMIndex}{\GaussianIndex}\in\R^{|\DimensionIndexSet|\times|\DimensionIndexSet|}$.
The density of a GMM at $\AttributePosition\in\R^{|\DimensionIndexSet|}$ is the weighted sum of the individual Gaussian densities (\autoref{fig:GaussiansGMM}):
$$\GaussianDensity{\GMMIndex}(\AttributePosition):=\sum_{\GaussianIndex=0}^{\GaussianCount{\GMMIndex}-1} \frac{\GaussianWeight{\GMMIndex}{\GaussianIndex}}{\sqrt{|2\pi\GaussianCovariance{\GMMIndex}{\GaussianIndex}|}}\exp\left(-\frac{\Transpose{(\AttributePosition-\GaussianMean{\GMMIndex}{\GaussianIndex})}\GaussianCovariance{\GMMIndex}{\GaussianIndex}^{-1}(\AttributePosition-\GaussianMean{\GMMIndex}{\GaussianIndex})}{2}\right).$$

We compute the parameters of a GMM from a sequence of input samples with the expectation maximization (EM) procedure. This iterative method seeks maximum likelihood estimates of the model parameters. It alternates between an expectation step, which evaluates the log-likelihood of the input samples using the current parameters, and a maximization step, which computes the parameters by maximizing the expected log-likelihood found in the expectation step.

\subsection{Fast Selection of GMM Components}
The EM algorithm takes the number of Gaussian components $\GaussianCount{\GMMIndex}$ as input. With more components, the target distribution can be modeled better. However, too many components may not significantly improve the model, but increase the storage overhead. A fixed, arbitrary number of number of components is often used~\cite{Dutta2017, Dutta2017a, Hazarika2019}. Similar to Wang et al.~\cite{Wang2017}, we adaptively select the appropriate number of components, but propose approximations to significantly reduce the computational complexity.

We iteratively fit GMMs with an increasing number of components up to a user specified maximum and select the GMM with the best Bayesian information criterion (BIC)~\cite{Schwarz1978}.
The BIC rewards a high likelihood over the training data and penalizes by the number of components. It is defined using the number of free parameters $\GMMFreeParams$ in the GMM as
$$
-2 \LogLikelihood + \GMMFreeParams \log|\ClusterIndexSet|,
$$
where $\LogLikelihood$ denotes the maximized log-likelihood and $\GMMFreeParams$ is given by
$$
\GMMFreeParams := \GaussianCount{\GMMIndex} \left( \frac{|\DimensionIndexSet|(|\DimensionIndexSet| + 1)}{2} + |\DimensionIndexSet| \right) + |\DimensionIndexSet| - 1.
$$

The iterative computation of GMMs with different numbers of components is computationally challenging, especially for large clusters. To speed up the selection of the best $\GaussianCount{\GMMIndex}$, we propose two approximations:
First, we take a random subset $\ClusterIndexSubSet \subset \ClusterIndexSet$ of our cluster whilst iteratively estimating the GMMs. After we have selected the best $\GaussianCount{\GMMIndex}$ based on the BIC, we recompute the GMM with $\GaussianCount{\GMMIndex}$ components for the whole cluster $\ClusterIndexSet$. 

Second, after we have selected the number of components $\{ \GaussianCount{0},\ldots,\GaussianCount{\DataAnyCount-1}\}$ for all one-dimensional GMMs, we use them as lower and upper bounds for the two- and three-dimensional GMMs. In detail, for a subset of attributes $\DimensionIndexSet\subseteq\{0,\ldots,\DataAnyCount-1\}$, we define the lower bound as
\begin{equation*}
\GaussianCount{\ClusterIndexSet,\DimensionIndexSet}^{\min} := \min_{\DataAnyIndex \in \DimensionIndexSet} \GaussianCount{\ClusterIndexSet,\{\DataAnyIndex\}}
\end{equation*}
and an approximate upper bound as
\begin{equation*}
\GaussianCount{\ClusterIndexSet,\DimensionIndexSet}^{\max} := \Pi_{\DataAnyIndex \in \DimensionIndexSet} \GaussianCount{\ClusterIndexSet, \{\DataAnyIndex\}}.
\end{equation*}
This implies that the higher-dimensional GMMs include at least the complexity of lower dimensions, whilst being bound by all combinations of all lower dimensional Gaussian components. In the supplementary material, we show that the bounds introduce no error, whilst the subsampling introduces a small error on our datasets.


Lastly, it is possible that some clusters contain only a small number of data samples. Although such a clustering may not seem optimal, it is quite likely to occur for scattered data.  For very small clusters, e.g.\ $|\ClusterIndexSet| \leq 20$, fitting a GMM is problematic since the target distribution may be underdetermined.
In this case, we fit a single Gaussian to these clusters.

\section{Spatial Visualization}
\label{sec:Spatial}

In this section, we discuss the visualization of the spatial density distribution. Although we could reconstruct the original data by drawing samples from the GMM of each cluster, this would require rendering a large amount of scattered data. Instead, we derive an efficient formulation to directly splat three-dimensional Gaussians. Additionally, we consider the application of a transfer function to a one-dimensional value distribution in each cluster.

\subsection{Integrating Visibility for Gaussians}

To render a trivariate Gaussian distribution, we integrate along a view ray $\RayOrigin+\RayVariable\RayDirection$ starting at $\RayOrigin\in\R^3$ in normalized direction $\RayDirection\in\R^3$ with $\RayVariable\in\R$. The Gaussian is given by its weight $\SingleGaussianWeight:=\GaussianWeight{\GMMIndex}{\GaussianIndex}$, mean $\SingleGaussianMean:=\GaussianMean{\GMMIndex}{\GaussianIndex}\in\R^3$ and covariance $\SingleGaussianCovariance:=\GaussianCovariance{\GMMIndex}{\GaussianIndex}\in\R^{3\times3}$. To derive a general solution, we integrate over $[\IntervalBegin, \IntervalEnd]$ by substituting the ray equation into the trivariate Gaussian distribution:
$$
\GaussianIntegral{\IntervalBegin}{\IntervalEnd} := \int_{\IntervalBegin}^{\IntervalEnd} \frac{1}{\sqrt{|2\pi\SingleGaussianCovariance|}} \exp \left(-\frac{\Transpose{(\RayOrigin + \RayVariable\RayDirection - \SingleGaussianMean)} \SingleGaussianCovariance^{-1} (\RayOrigin + \RayVariable\RayDirection - \SingleGaussianMean)}{2} \right)\,\textrm{d}\RayVariable.
$$
Through integration by substitution (see the supplementary material), we obtain the following closed-form solution:
\begin{equation}
\label{eq:Gaussian}
\GaussianIntegral{\IntervalBegin}{\IntervalEnd} = \IntegralAuxiliary\frac{\sqrt{\pi}}{\sqrt{\IntegralAuxiliaryDirection}} \left[\frac{1}{2} \erf(\TransformedRayVariable)\right]_{\sqrt{\IntegralAuxiliaryDirection} \left(\IntervalBegin + \frac{\IntegralAuxiliaryOffsetDirection}{\IntegralAuxiliaryDirection}\right)}^{\sqrt{\IntegralAuxiliaryDirection} \left(\IntervalEnd + \frac{\IntegralAuxiliaryOffsetDirection}{\IntegralAuxiliaryDirection}\right)},
\end{equation}
with
\begin{align*}
\IntegralAuxiliaryDirection &:= \frac{1}{2} \Transpose{\RayDirection} \SingleGaussianCovariance^{-1} \RayDirection \text{,} \\
\IntegralAuxiliaryOffsetDirection &:= \frac{1}{2} \Transpose{(\RayOrigin-\SingleGaussianMean)} \SingleGaussianCovariance^{-1} \RayDirection\text{,} \\
\IntegralAuxiliary &:= \frac{1}{\sqrt{|2\pi\SingleGaussianCovariance|}} \exp\left(-\frac{1}{2} \Transpose{(\RayOrigin-\SingleGaussianMean)} \SingleGaussianCovariance^{-1} (\RayOrigin-\SingleGaussianMean) + \frac{\IntegralAuxiliaryOffsetDirection^2}{\IntegralAuxiliaryDirection}\right) \text{.}
\end{align*}
When integrating over all of $\R$, this result simplifies to
\begin{equation}
\label{eq:GaussianInfty}
\GaussianIntegral{-\infty}{\infty} = \IntegralAuxiliary \frac{\sqrt{\pi}}{\sqrt{\IntegralAuxiliaryDirection}}.
\end{equation}
\revi{We could use this result inside a ray tracer, possibly with ray tracing GPUs~\cite{Knoll2019}. It only has to identify relevant Gaussians per pixel, ray marching for integration becomes unnecessary. In the following, we discuss our approach using GPU rasterization, which works efficiently on commodity graphics hardware.}

\subsection{Back-to-Front Splatting}
To splat scattered 3D Gaussians, we sort them from back-to-front based on their mean distance to the camera.
Then, we integrate the Gaussians along the viewing direction using \autoref{eq:GaussianInfty}. Integrating from $-\infty$ to $\infty$ is generally a reasonable approximation, but it is possible that we incorrectly evaluate a Gaussian if the camera is positioned within its support. Alternatively, we could employ \autoref{eq:Gaussian}, but this is far more expensive and only gives a benefit in rare cases.

To render a single 3D Gaussian, we first compute the principal components of the distribution to fit a bounding box along the principal axes. By default, we limit the size of the bounding box in each dimension by 3 standard deviations.
This box is then rasterized and for each resulting fragment, we integrate the Gaussian along the viewing direction in a fragment shader using \autoref{eq:GaussianInfty}. Finally, we tone-map the resulting density (see \autoref{eq:ToneMapping}) to better convey the high-dynamic range.

We did experience some numerical issues with some of our datasets due to very large or small spatial and value domains. We were able to address these issues by switching to a more numerically stable eigen decomposition~\cite[Algorithm 8.2.3]{Golub1993}. Lastly, we make use of the Cholesky decomposition to invert covariance matrices, which behaves robustly even for nearly singular matrices~\cite[p. 176]{Trefethen1997}.

\subsection{Order-Independent Transparency}

The splatting approach assumes that distributions do not overlap since this could lead to visible flickering between frames when the order changes. 
Depending on the clustering of the data, this assumption is not always acceptable.
\revi{For this reason, we propose the use of an order-independent transparency (OIT) approach to avoid sorting semi-transparent Gaussians. Although, a large number of Gaussians are problematic for most OIT approaches,} moment-based order-independent transparency (MBOIT)~\cite{Muenstermann2018} is well-suited for this application. We introduce the steps of this method briefly.

MBOIT first accumulates moments of the optical depth in an additive rendering pass. These moments offer a compact representation and an efficient reconstruction of the transmittance function per view ray. Subsequently, a second additive rendering pass of all Gaussians composites the fragment colors using transmittance values reconstructed from the moments. We use three trigonometric moments in half-precision, which results in a total of 112 bits per pixel for the moments.


\subsection{Uncertainty Transfer Function}

Lastly, we discuss how to apply a transfer function to the distribution of an attribute $\DataAnyIndex \in \{ \GaussianCount{0},\ldots,\GaussianCount{\DataAnyCount-1}\}$ \revirm{to obtain a color and opacity }for each cluster $\ClusterIndexSet$. The one-dimensional value dimension is modeled \revi{separately from the cluster} as a GMM $\GMMIndex = (\ClusterIndexSet, \{\DataAnyIndex \})$ with $\GaussianCount{\GMMIndex}$ components. \revi{For each cluster, we compute an expected color and opacity~\cite{Sakhaee2017} by convolving the transfer function $\TF$ with the value distribution:}
%
%
\begin{equation*}
E[\TF | \GMMIndex ] := \int_{-\infty}^{\infty} \TF(\AttributePosition) \GaussianDensity{\GMMIndex}(\AttributePosition) \textrm{d}\AttributePosition.
\end{equation*}
We insert the Gaussian mixture model into this equation and rearrange:
\begin{equation}
\label{eq:TransferFuncExpectation}
E[\TF | \GMMIndex ] = \sum_{\GaussianIndex=0}^{\GaussianCount{\GMMIndex}-1} \GaussianWeight{\GMMIndex}{\GaussianIndex} \int_{-\infty}^{\infty} \TF(\AttributePosition)
\frac{1}{\sqrt{2\pi \GaussianOneDimVariance{\GMMIndex}{\GaussianIndex}}} \exp \left(-  \frac{(\AttributePosition - \GaussianOneDimMean{\GMMIndex}{\GaussianIndex})^2}{2 \GaussianOneDimVariance{\GMMIndex}{\GaussianIndex}} \right)
\textrm{d}\AttributePosition.
\end{equation}
We efficiently evaluate this equation by precomputing the integrand, which is simply a convolution of the transfer function with differently parametrized Gaussians. The resulting 2D lookup table thus depends on the transfer function and is parameterized by mean and variance. 

\section{Visual Analysis}
\label{sec:IVA}

Now that we are able to render the spatial distribution of our data, we move on to the visual exploration and analysis of additional data dimensions using our representation. This includes multiple views with brushing and linking coupled with a focus and context visualization to emphasize brushed values.

\ifdefined\ShowRevision
\subsection{\revirm{Sample-Based Visualization}}
\fi

%

\subsection{Density-Based Visualization}

\revi{
Prior work relies on sampling to create visualizations from modeled distributions. Although this is similarly possible with our data representation, see \autoref{fig:SprayNozzleSamples} (c) and (d), we focus on density-based visualizations.}
Since we already have an estimate of density in the form of our GMMs, we efficiently construct density-based visualizations that are costly to compute otherwise. To obtain the density, we evaluate and accumulate the Gaussian distributions from the GMMs in all clusters.
Since the distribution in each cluster is normalized, we additionally weight each cluster $\ClusterIndexSet$ by the normalized number of samples it represents $\frac{1}{\SampleCount}|\ClusterIndexSet|$. 

\subsubsection{Density Plots}

To compute a density in 1D or 2D, we evaluate and accumulate the Gaussian distributions, see \autoref{fig:teaser}(b) and (c). Since this operation can be parallelized trivially, we make use of GPU acceleration. In the 1D case, we evaluate 1D Gaussians on the GPU and plot a probability density function. For a 2D plot, we render a quadrilateral for each Gaussian, evaluate the Gaussian for each fragment, and additively accumulate the results.

\subsubsection{Parallel Coordinate Plots}

Miller and Wegman~\cite{Miller1991} formulate parallel coordinate plots for bivariate Gaussian distributions.
With this formulation, we splat the 2D distributions for each pair of consecutive dimensions in parallel coordinate space. Specifically, for each pair of axes we draw a quad for each Gaussian by truncating its support to three standard deviations. In the fragment shader, we evaluate the density in parallel coordinate space and additively blend the result with all other Gaussians. A density-based parallel coordinate plot is shown in \autoref{fig:teaser}(d).

\subsubsection{Mapping Density}
The density-based visualizations described above and the spatial visualization in \autoref{sec:Spatial} all produce a single density $\Density$ per Gaussian and per pixel.
For large datasets, this density will have a high dynamic range and needs to be mapped to an opacity between zero and one through a non-linear mapping~\cite{Johansson2005}.
We choose a mapping that interprets the density, scaled by a user-controllable parameter $\DensityScale>0$, as optical depth. The resulting opacity is
\begin{equation}
\label{eq:ToneMapping}
1 - \exp(-\DensityScale\Density).
\end{equation}
With this mapping, multiplying the density of a Gaussian by an integer factor $k$ produces the same result as rendering it $k$ times with alpha blending, which is an intuitive behavior. At the same time, it retains detail even for large densities.

\subsection{Brushing Distributions}
\label{sec:IVA:Brushing}

\begin{figure}
	\centering
	\subfloat[Brushing (mean)]{%
		\includegraphics[width=0.325\linewidth]{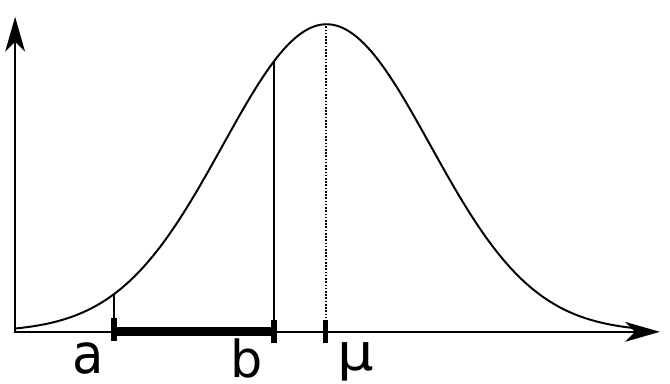}
	}
	\subfloat[Brushing (mean)]{%
		\includegraphics[width=0.325\linewidth]{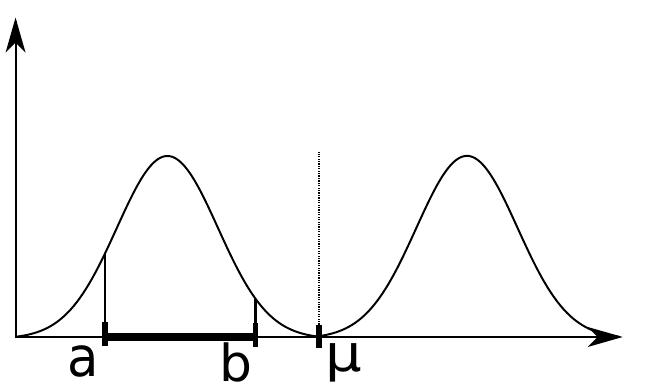}
	}
	\subfloat[Brushing (distribution)]{%
	\includegraphics[width=0.325\linewidth]{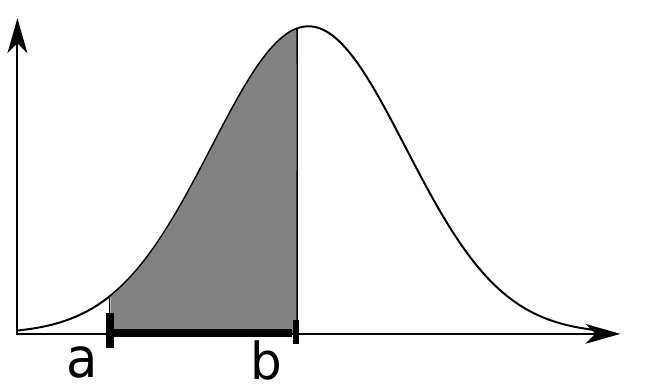}
	}
	\vspace{-1mm}
	\caption{Brushing of a value range $[\IntervalBegin,\IntervalEnd]$ applied to several distributions. Brushing only based on the mean value $\SingleGaussianMean$ would lead to confusing results (a), especially if the distribution is represented by multiple Gaussian components (b). We compute the degree of interest of the brushing operation as the ratio between the integrand (gray) and the total area under the curve (c).}
	\label{fig:SmoothBrushing}
	\vspace{-3mm}
\end{figure}

To brush a value range of a dimension with our data representation and reflect this in all linked views, we use the clustering information. Although we could brush based on the cluster mean value, this is confusing and not very intuitive, especially when considering a dimension represented by multiple Gaussian components, see \autoref{fig:SmoothBrushing}(a) and (b).

Feng et al.~\cite{Feng2010} \revi{discuss} user interaction based on \revi{Gaussian} distributions in the context of uncertainty visualization.
\revi{We generalize their work and the concept of smooth brushing~\cite{Doleisch2001} to Gaussian mixture models. In detail,} we compute the amount a cluster is in focus, the degree of interest, as the ratio between the integrand of the brushed regions of the \revi{GMM} and the total area, see \autoref{fig:SmoothBrushing}(c). For clusters that contain multiple Gaussian components, we compute the degree of interest as the weighted sum of all components.

\subsection{Time-Dependent Visualization}
\label{sec:IVA:Time}

Brushing in different time steps is a powerful tool for the interactive exploration of time-dependent data \cite{Hochheiser2004}, but is not practical for large datasets since all time steps have to be processed. Our compact data summaries enable us to interact with multiple time steps at once. We support this interaction in a time histogram~\cite{Kosara2004} where we depict a time-series of a selected dimension as a series of 1D histograms, see~\autoref{fig:Illustris3_TimeHist_Details}.

If the clustering is fixed over time, we can trivially extend the brushing operation to time-dependent data. This is not possible when clusters change over time, e.g.\ merge together into larger, or split into smaller clusters. In this case, the relationship of clusters in different time steps has to be explicitly modeled and stored.

For brushing, we need to reassign degrees of interest from frame to frame. To this end, we transfer the degrees of interest to the individual samples uniformly and then reassign them to clusters. Say we have $\ClusterCountCurrent\in\N$ clusters $\ClusterIndexSetCurrent{0},\ldots,\ClusterIndexSetCurrent{\ClusterCountCurrent-1}\subseteq\{0,\ldots,\SampleCount-1\}$ in frame $\FrameIndex$ and analogously for frame $\FrameIndex+1$. The clusters in frame $\FrameIndex$ have associated degrees of interest $\ClusterDegreeOfInterestCurrent{0},\ldots,\ClusterDegreeOfInterestCurrent{\ClusterCountCurrent-1}\in[0,1]$.
Then we define the degree of interest of cluster $\ClusterIndex\in\{0,\ldots,\ClusterCountNext-1\}$ in frame $\FrameIndex+1$ as
\begin{equation*}
\ClusterDegreeOfInterestNext{\ClusterIndex}:=\sum_{\OtherClusterIndex=0}^{\ClusterCountCurrent-1}\frac{|\ClusterIndexSetCurrent{\OtherClusterIndex}\cap\ClusterIndexSetNext{\ClusterIndex}|}{|\ClusterIndexSetNext{\ClusterIndex}|}\ClusterDegreeOfInterestCurrent{\OtherClusterIndex}\in[0,1]\text{.}
\end{equation*}
The quotient in this sum is the fraction of samples in cluster $\ClusterIndexSetNext{\ClusterIndex}$ that was part of cluster $\ClusterIndexSetCurrent{\OtherClusterIndex}$ in the previous frame. Interest is inherited from the cluster in the previous frame in proportion to that quotient.
Note that this method defines a simple linear transform. There is no need to consider all samples at run time. Instead, the transfer coefficients for the degrees of interest can be precomputed and stored in a sparse matrix.

\subsection{Uncertainty Visualization}
\label{sec:IVA:Error}

We introduce an error estimate to convey the uncertainty of the data summaries.
By computing and storing an error for each cluster, we are able to visualize the uncertainty interactively during the visual analysis and to support brushing and linking.
\revi{Prior work measures the error directly between the density of the Gaussian mixture model and the original data. However, this is not robust and suffers from aliasing due to the necessary use of histograms. Instead, we define t}he error between a Gaussian mixture model and the samples of a cluster $\ClusterIndexSet$ for a dimension $\DataAnyIndex\in\{0,\ldots,\DataAnyCount-1\}$ \revirm{is defined }similar to common statistical goodness of fit tests. In detail, we compute the empirical cumulative distribution functions (CDF) of the data samples
\begin{equation}
\CdfDataSamples(\AttributePosition) := \frac{1}{|\ClusterIndexSet|} \sum_{\SampleIndex \in \ClusterIndexSet}
\begin{cases}
1 & \text{if } \DataAny{\SampleIndex}{\DataAnyIndex} \leq \AttributePosition, \\
0 & \text{otherwise}
\end{cases}
\end{equation}
and compare it to the CDF $\CdfGMM$ of the Gaussian mixture model using the Wasserstein distance~\cite{Wasserstein2019}:
\begin{equation}
\Wasserstein(\CdfDataSamples, \CdfGMM) := \int_{-\infty}^{\infty} \left| \CdfDataSamples(\AttributePosition) - \CdfGMM(\AttributePosition)\right| \textrm{d}\AttributePosition.
\end{equation}
To visualize the Wasserstein distance, we show it together with the CDF, cf.~\autoref{fig:synth:error} (b). A high Wasserstein distance consequently indicates a high uncertainty of the data model.\revirm{ In contrast, measuring an error from the probability density function is not robust and suffers from aliasing due to the necessary use of histograms.}

\subsection{Level of Detail and Outliers}
\label{sec:IVA:LOD}

\begin{figure}
	\centering
	\includegraphics[width=\linewidth]{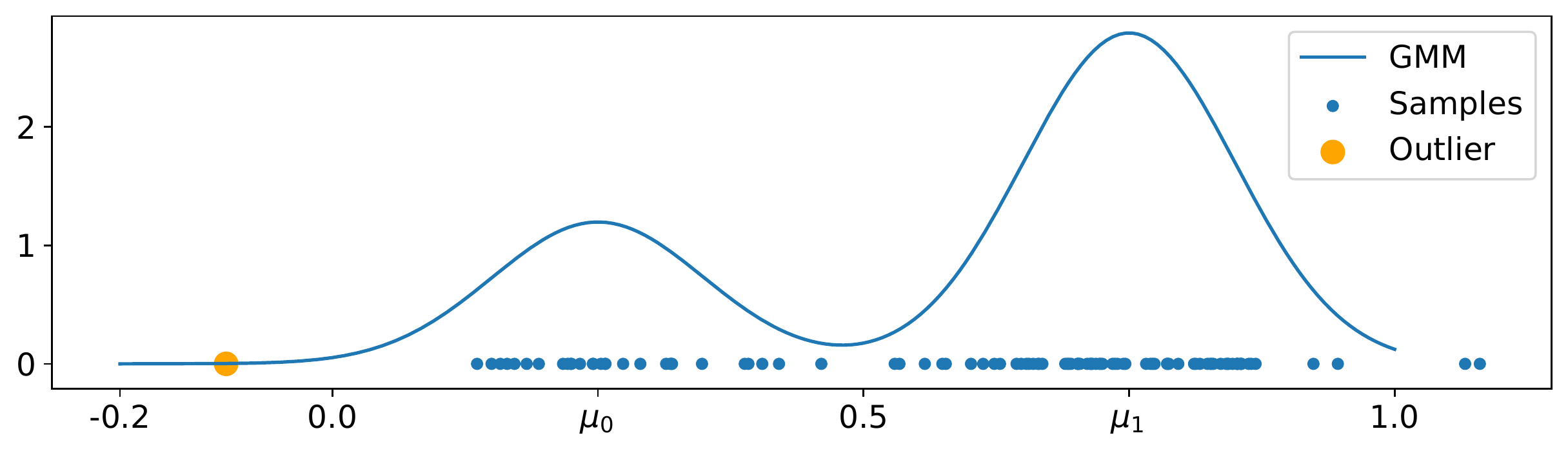}
	\vspace{-5mm}
	\caption{To rank samples by their outlyingness, we evaluate the Mahalanobis distance to the closest Gaussian. This measures how many standard deviations a sample is away from the mean of the closest Gaussian.}
	\label{fig:Outlier}
	\vspace{-2mm}
\end{figure}

By design, our representation is a simplified model of the data. During the exploration and analysis process, a scientist might want to investigate a subset of the data more closely. For this purpose, we substitute brushed clusters by their original data values.
To integrate the data distributions into our frequency-based views, we perform kernel density estimation using Gaussian kernels. We can thus avoid differentiating between the modeled and original data distributions.

Moreover, outliers, i.e.\ isolated samples in regions of low density, tend to get lost in density-based visualizations~\cite{Novotny2006, Mayorga2013}.
To explicitly add outliers to our visualizations, we sort all samples in a cluster in a preprocess according to a measure of outlyingness. Although any measure between a sample and a GMM could be used, we employ the Mahalanobis distance~\cite{Mahalanobis1936} to the closest Gaussian component. This effectively measures how many standard deviations a sample is away from the mean of the closest Gaussian, see \autoref{fig:Outlier}.
To visualize outliers, we then take a fixed percentage $\OutlierPercentage$ of outliers from a cluster $\ClusterIndexSet$ by loading the first $\OutlierPercentage |\ClusterIndexSet|$ samples. \autoref{fig:synth:spatial} (c) shows a spatial visualization with \num{2}\si{\percent} of outliers.


\section{Evaluation}
\label{sec:Results}

In this section, we apply our approach to a synthetic and three real-world datasets. Additional results can be found in the supplementary material.

\subsection{Synthetic Data}
\label{sec:Results:Synth}
\begin{figure}
	\centering
	\subfloat[Synthetic data]{%
		\includegraphics[width=0.324\linewidth]{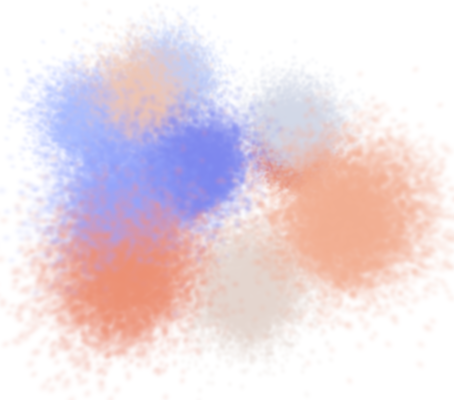}
	}
	\hfill
	\subfloat[GMMs]{%
		\includegraphics[width=0.324\linewidth]{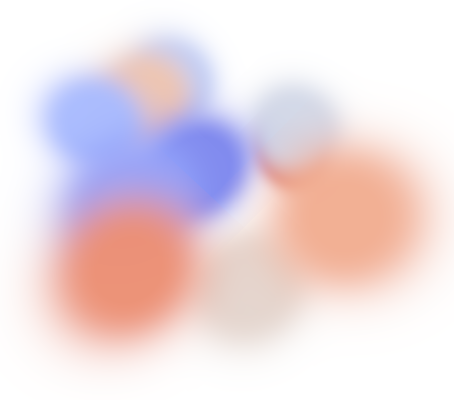}
	}
	\hfill
	\subfloat[GMMs and outliers]{%
		\includegraphics[width=0.324\linewidth]{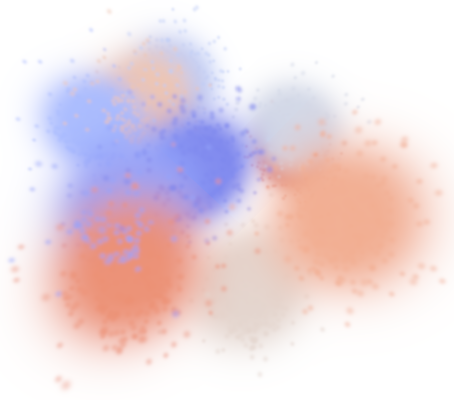}
	}
	\vspace{-1mm}
	\caption{Rendering of Gaussians from the synthetic dataset using kernel density estimation (a), with our data model (b), and with $2\%$ of outliers (c).
	}
	\label{fig:synth:spatial}
	\vspace{-3mm}
\end{figure}

\begin{figure}
	\centering
	\subfloat[Transfer function]{%
		\includegraphics[width=0.98\linewidth]{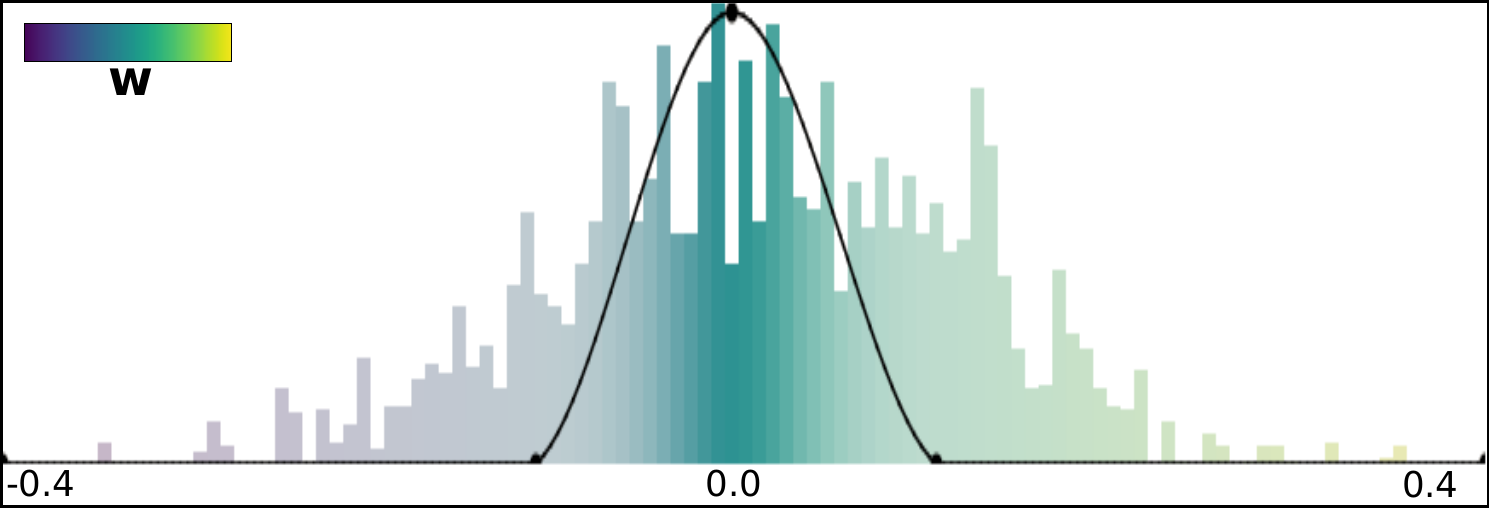}
	}
	\\
	\vspace*{-3mm}
	\subfloat[Synthetic data]{%
		\includegraphics[width=0.32\linewidth]{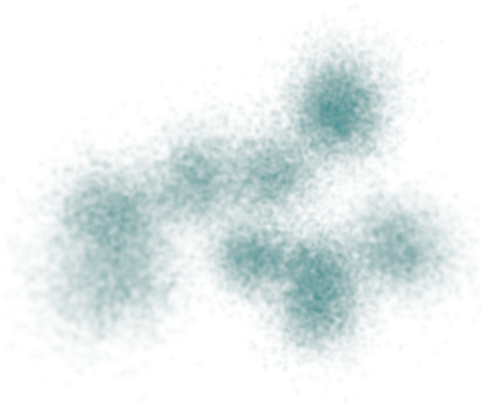}
	}
	\hfill
	\subfloat[Expected opacity]{%
		\includegraphics[width=0.32\linewidth]{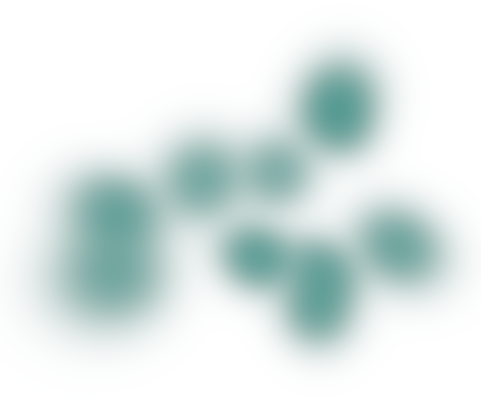}
	}
	\hfill
	\subfloat[Mean opacity]{%
		\includegraphics[width=0.32\linewidth]{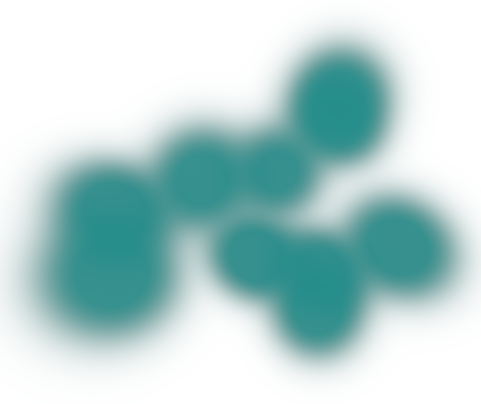}
	}
	\vspace{-1mm}
	\caption{We set the opacity of the transfer function (a) to visualize the synthetic dataset (b). Our uncertainty transfer function (c) computes the expected opacity (i.e.\ the integral of the opacity curve), while a 1D transfer function based on the mean value (d) sets all Gaussians to opaque.
	}
	\label{fig:synth:utf}
	\vspace{-2mm}
\end{figure}

We first apply our approach to a small synthetic dataset consisting of \num{10} clusters from a total of \num{100000} points. The dataset contains 9 dimensions. 
The three spatial dimensions in each cluster are normally distributed, but \SI{10}{\percent} of the points are distributed uniformly to add noise to the distributions. \autoref{fig:synth:spatial} (a) shows this dataset.
The 3D Gaussians are shown in \autoref{fig:synth:spatial} (b) and in (c) where we explicitly add $2\%$ of outliers from all clusters.

We compare the uncertainty transfer function to a 1D transfer function based on mean values in \autoref{fig:synth:utf}. In (a), we set the opacity of the transfer function, where the peak coincides with the mean value. The synthetic dataset in (b) shows the resulting rendering. For our data summaries in (c), the opacity is similarly reduced. This is due to the uncertainty transfer function since it computes the expected opacity with respect to the value distribution. In comparison, the opacity of the Gaussians in (d) using a mean transfer function does not change since the opacity of the mean value is still set to opaque in (a). Thus, changing any of the opacity (or color) values of the transfer function has no influence except if the mean value is changed. \revi{An alternative would be the use of a 2D transfer function~\cite{Kniss2002} that offers increased control over the classification, but complicates user interaction.}

\begin{figure}
	\centering
	\subfloat[Probability density function]{%
		\includegraphics[width=0.49\linewidth]{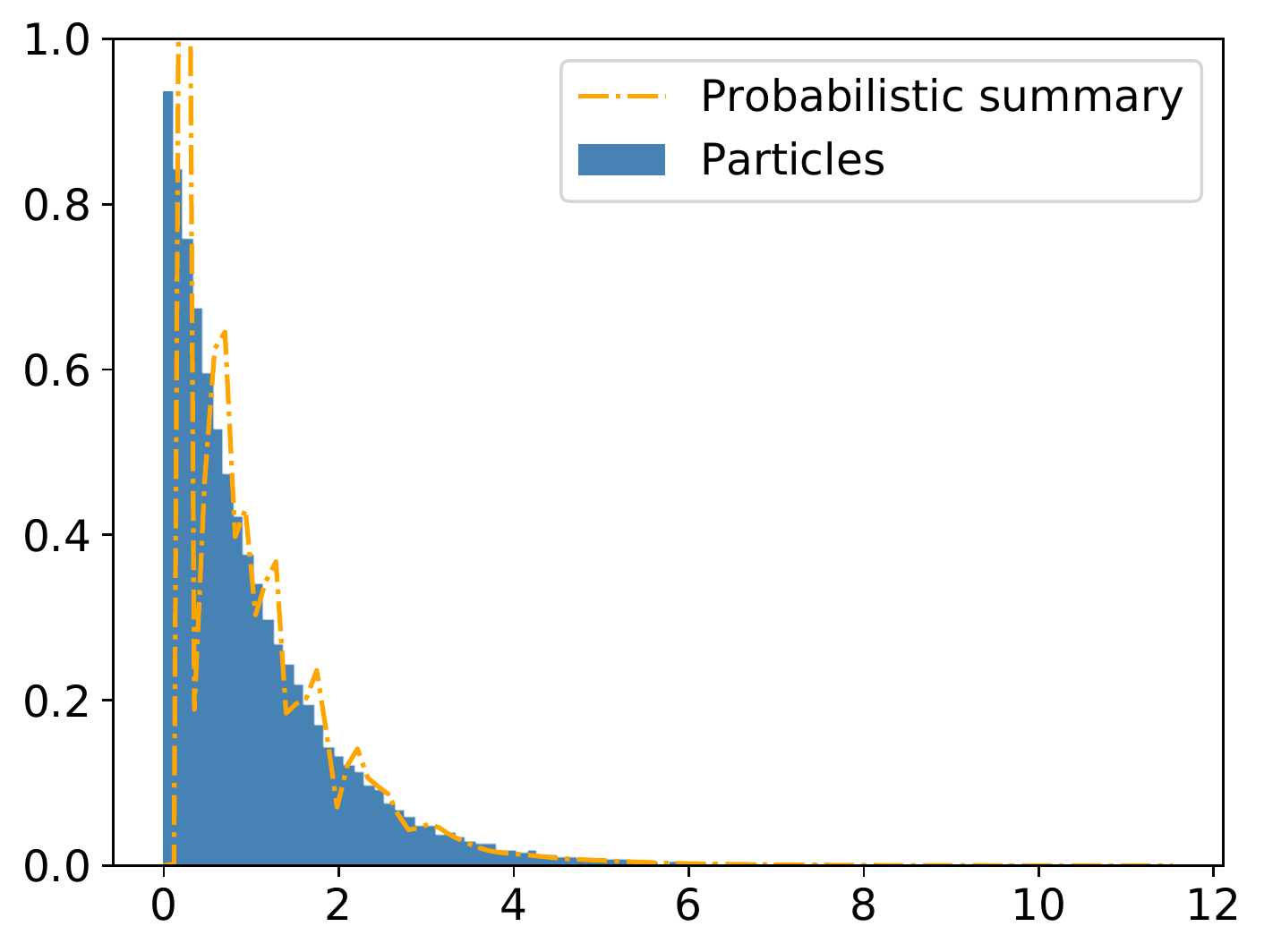}
	}
	\hfill
	\subfloat[Cumulative distribution function]{%
		\includegraphics[width=0.49\linewidth]{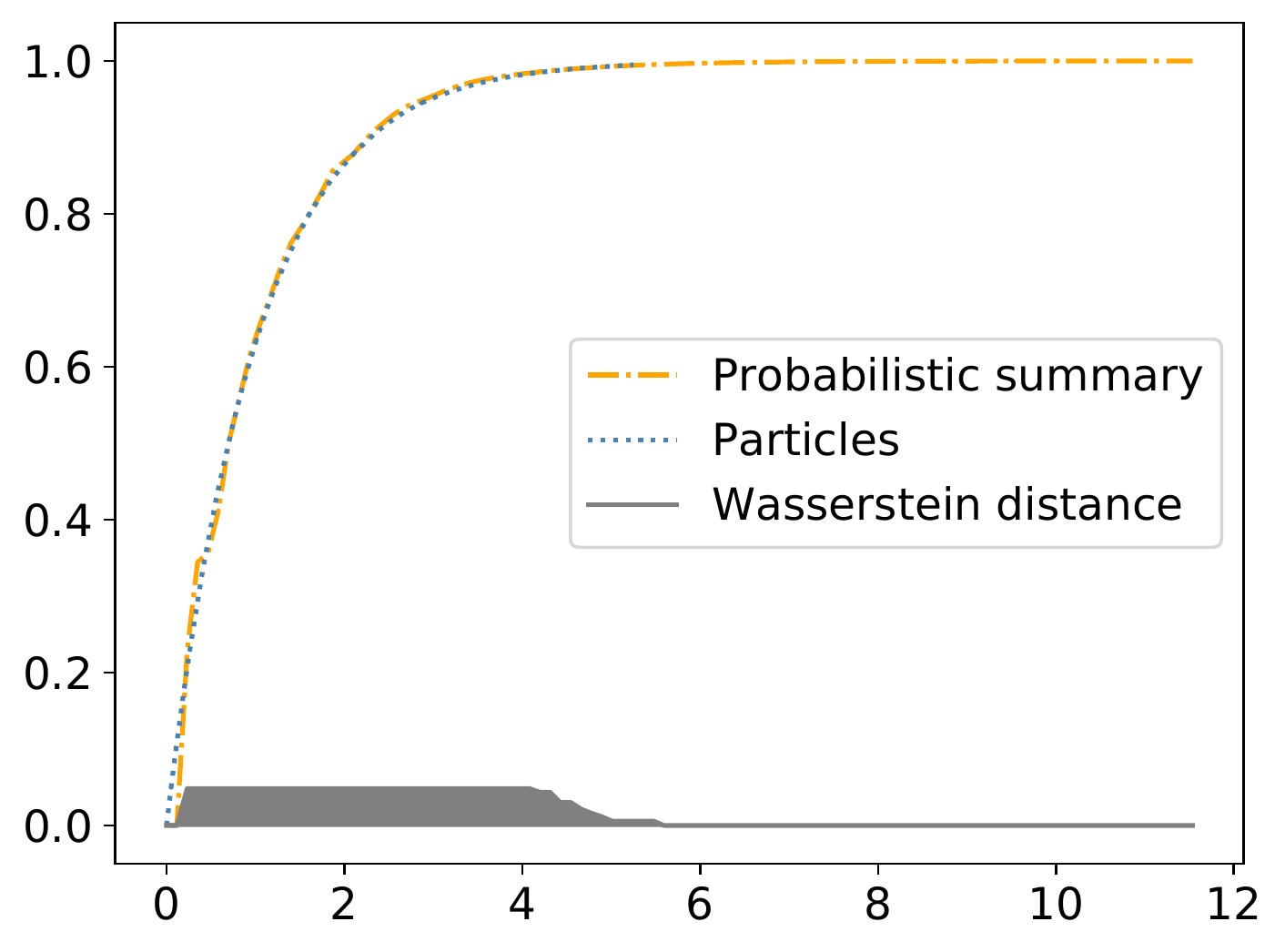}
	}
	\vspace{-1mm}
	\caption{
		The exponentially distributed dimension (a) is hard to model using Gaussian components. The cumulative distribution function in (b) conveys the error to the user. 
	}
	\label{fig:synth:error}
	\vspace{-3mm}
\end{figure}

In \autoref{fig:synth:error} (a) we illustrate an exponentially distributed dimension of the dataset, which is difficult to model using only Gaussian components. The cumulative distribution function shown in (b), illustrates this error as measured by the Wasserstein distance. By quantifying the error, we can decide if this error is acceptable, or brush and use the level-of-detail mechanism to directly load a subset of the data with a high error.

\subsection{Cosmological Data}
\label{sec:Results:Illustris}

\begin{table*}
	\centering
	\caption{Overview of the cosmological data from the Illustris simulation. }
	\vspace{-1mm}
	\label{tbl:IllustrisData}
		\begin{tabular}{lccccccc}
			\toprule
			Dataset & \# Dimensions & \# Particles & \# Clusters & Data size & Summaries & GMM comp. & Wasserstein dist. \\
			\midrule
			Illustris-3 Gas & 15 & 16,039,182 & 110,000 & 2 GB & 200 MB & $2.15 \pm 1.60$ & \num{1.77e-6}\\
			Illustris-2 DM & 7 & 319,324,195 & 841,639 & 11.3 GB & 617 MB & $1.54 \pm 1.22$ & \num{3.10e-07}\\
			Illustris-1 DM & 6 & 2,635,739,426 & 5,352,571 & 72.2 GB & 1.5 GB & $1.13 \pm 0.38$ & \num{4.56e-8}\\
			\bottomrule
	\end{tabular}
\end{table*}

\begin{figure*}
	\centering
	\includegraphics[width=\linewidth]{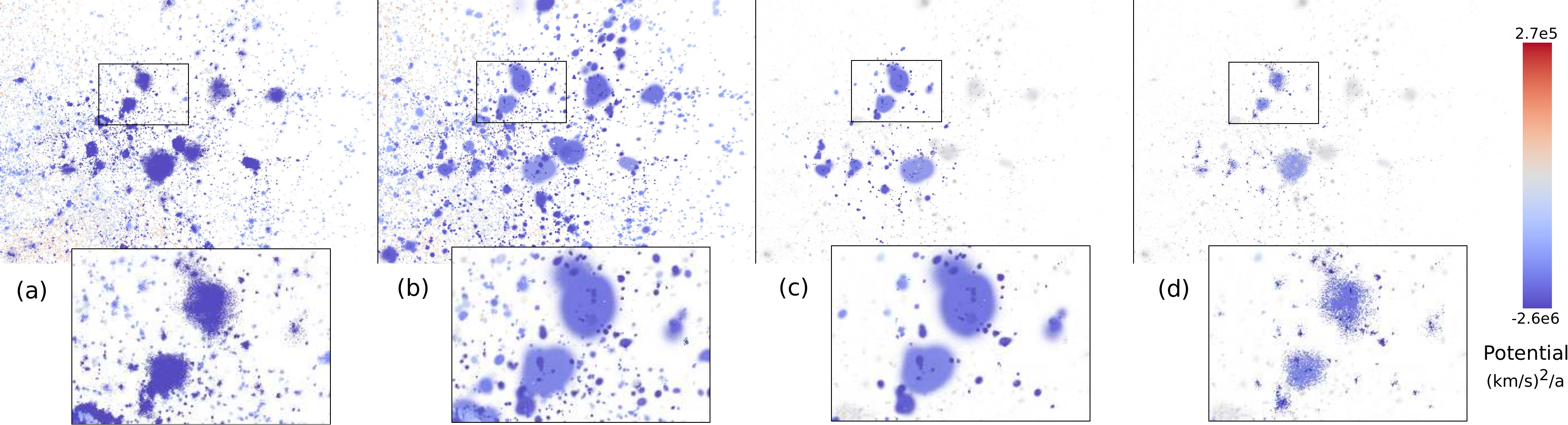}
	\caption{The Illustris-3 Gas dataset rendered by splatting particles (a) and 3D Gaussians (b). In (c) we have brushed a region and clusters that are not in focus are shown in a desaturated gray. We load the original particle data of the brushed region and render them together with the context (d).}
	\label{fig:Illustris3_Comparison}
\end{figure*}

\begin{figure*}
	\centering
	\subfloat[Time histogram]{%
		\includegraphics[height=42mm]{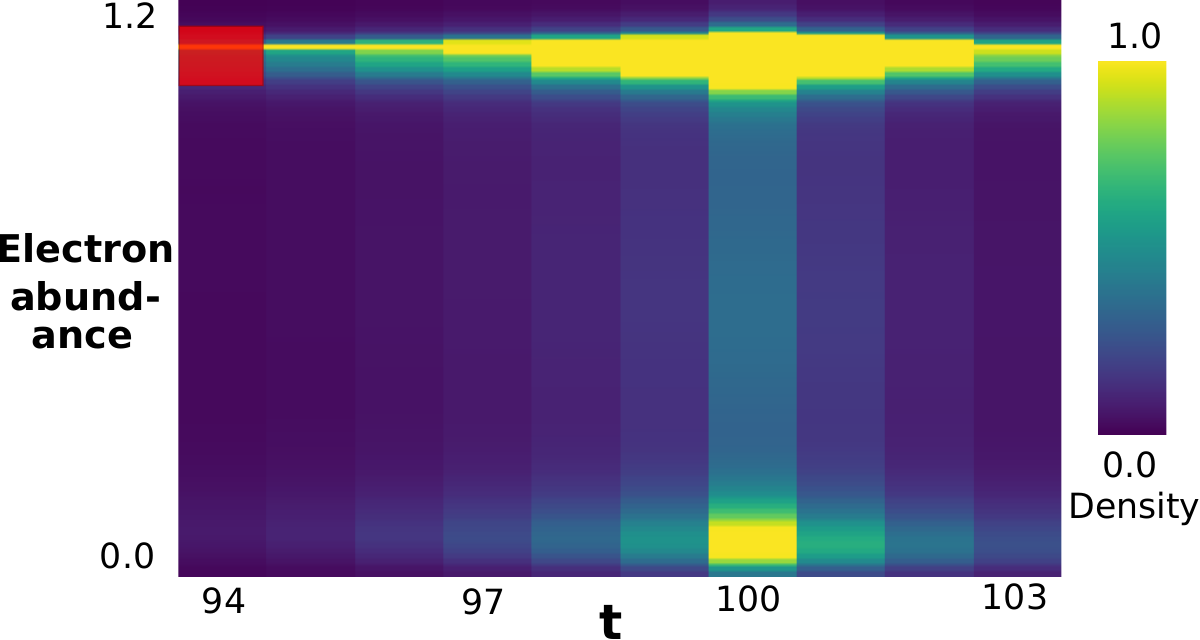}
	}
	\hfill
	\subfloat[Brushed clusters in the 100th time step]{%
		\includegraphics[height=45mm]{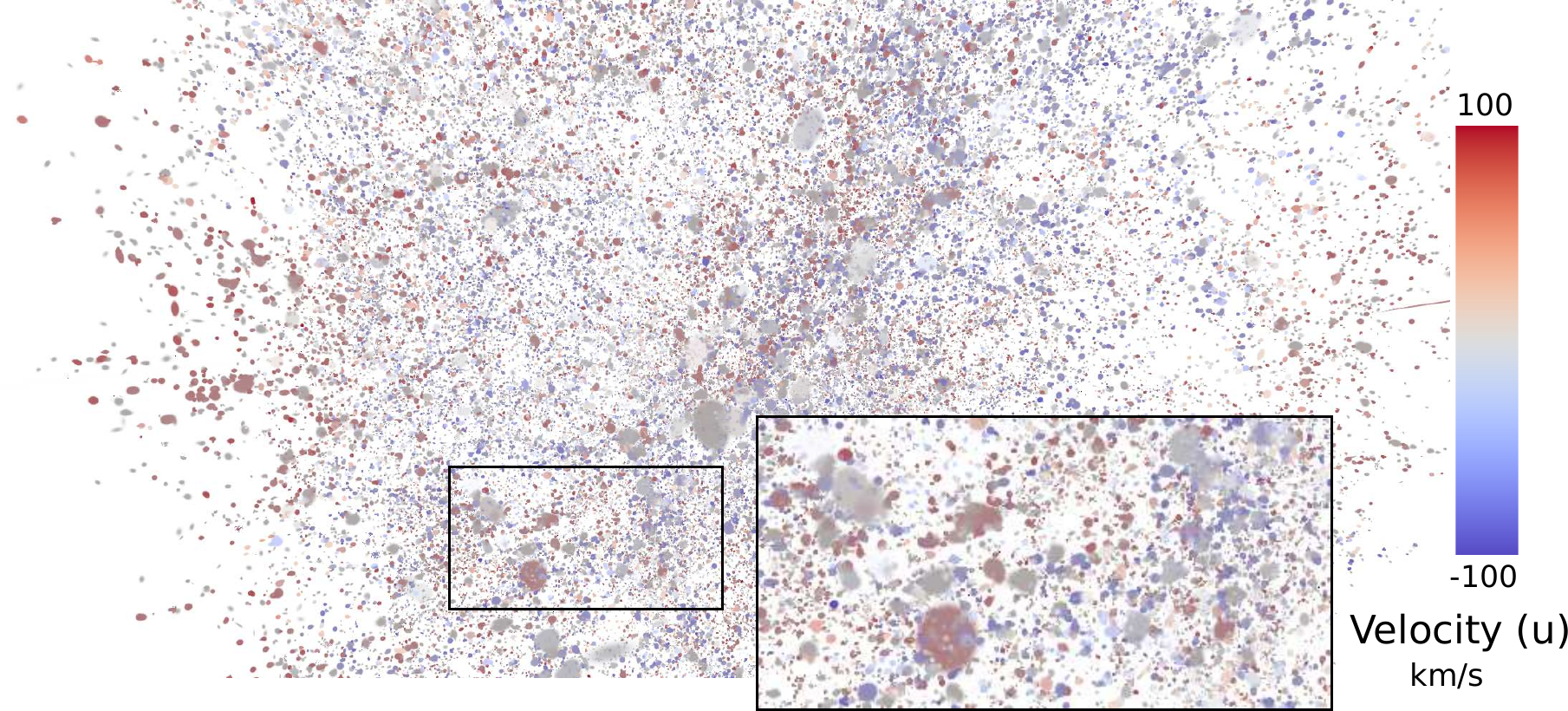}
	}
	\vspace*{-1mm}
	\caption{A time-histogram of electron abundance in the Illustris-3 Gas dataset is shown in (a). We have brushed (red) in the 94th time step, which affects all linked views in the current, 100th time step. The spatial visualization that highlights the brushed values in the 100th time step is shown in (b). Note that Gaussians are shown as saturated or desaturated, depending on how much they are in focus.}
	\label{fig:Illustris3_TimeHist_Details}
\end{figure*}

\begin{figure*}
	\centering
	\subfloat[Splatting 3D Gaussians ($k$-means 32.000)]{%
		\includegraphics[width=0.49\linewidth]{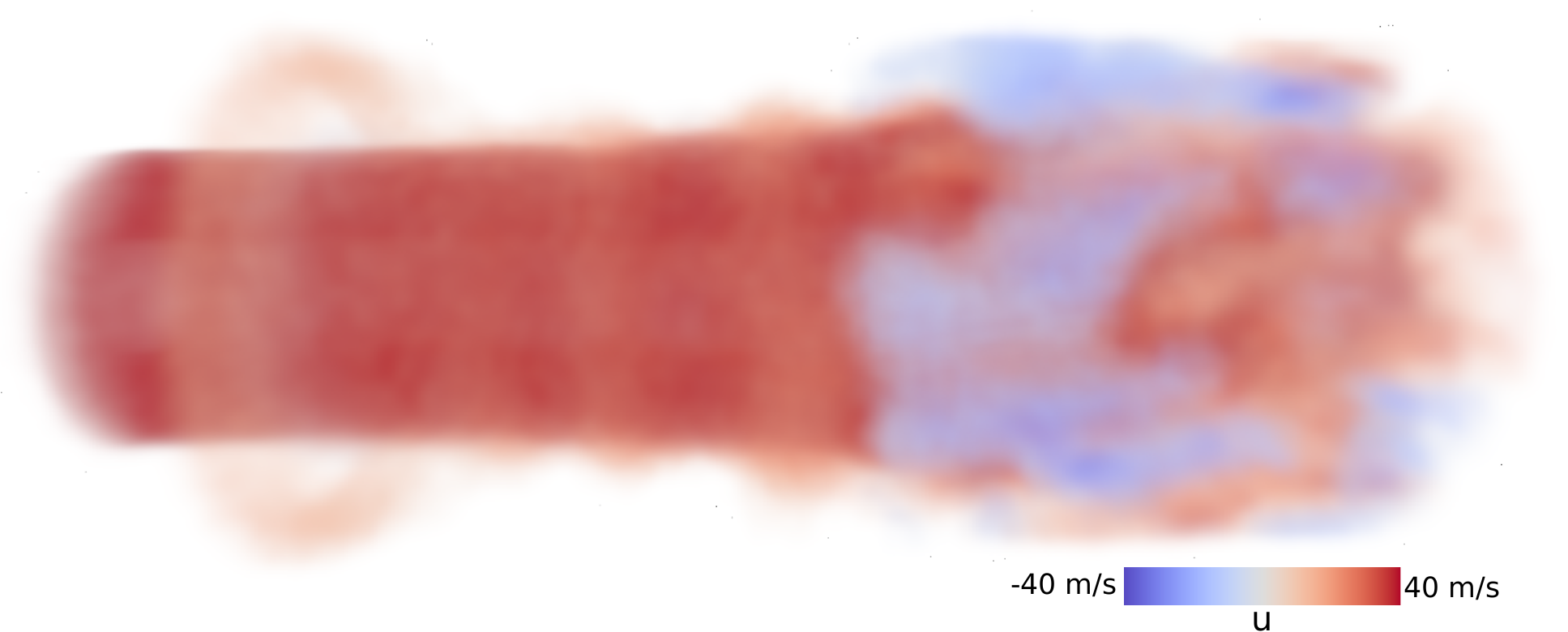}
	}
	\hfill
	\subfloat[Splatting all particles]{%
		\includegraphics[width=0.49\linewidth]{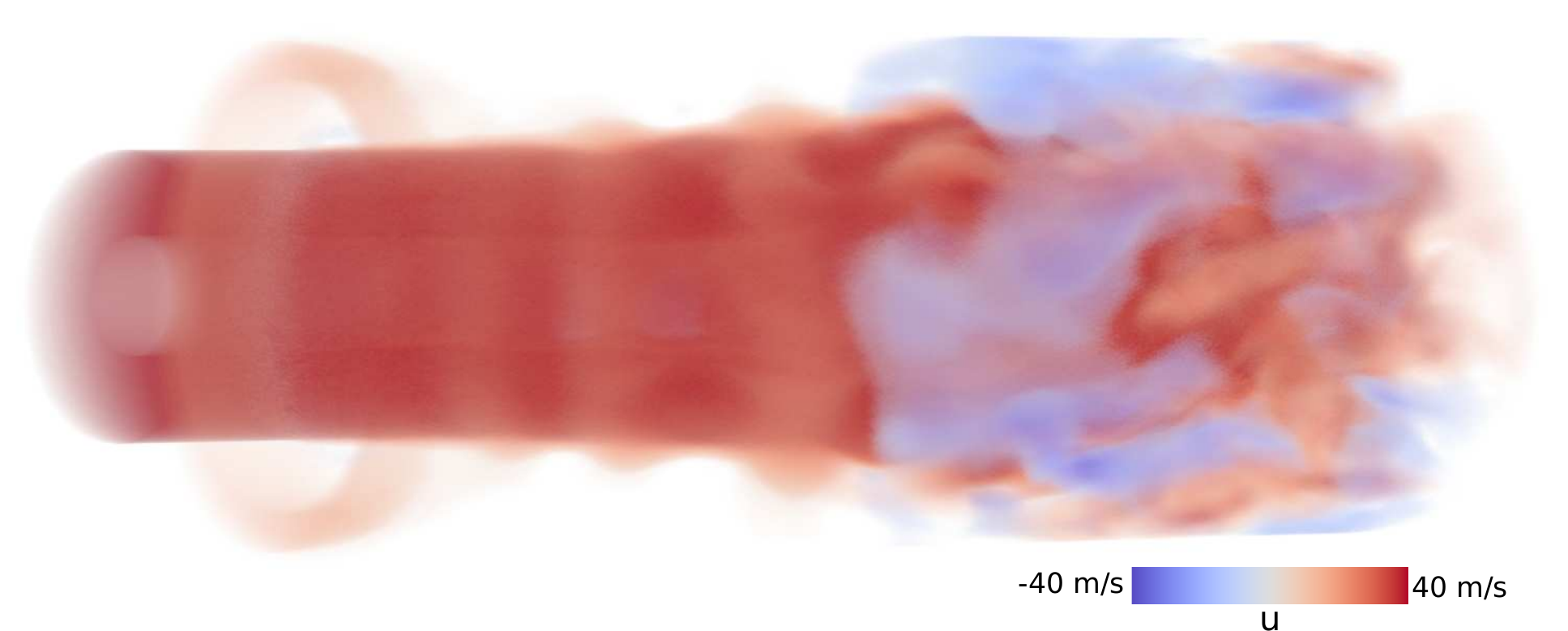}
	}
	\\
	\vspace{-1mm}
	\subfloat[Histogram from samples]{%
		\includegraphics[height=22.5mm]{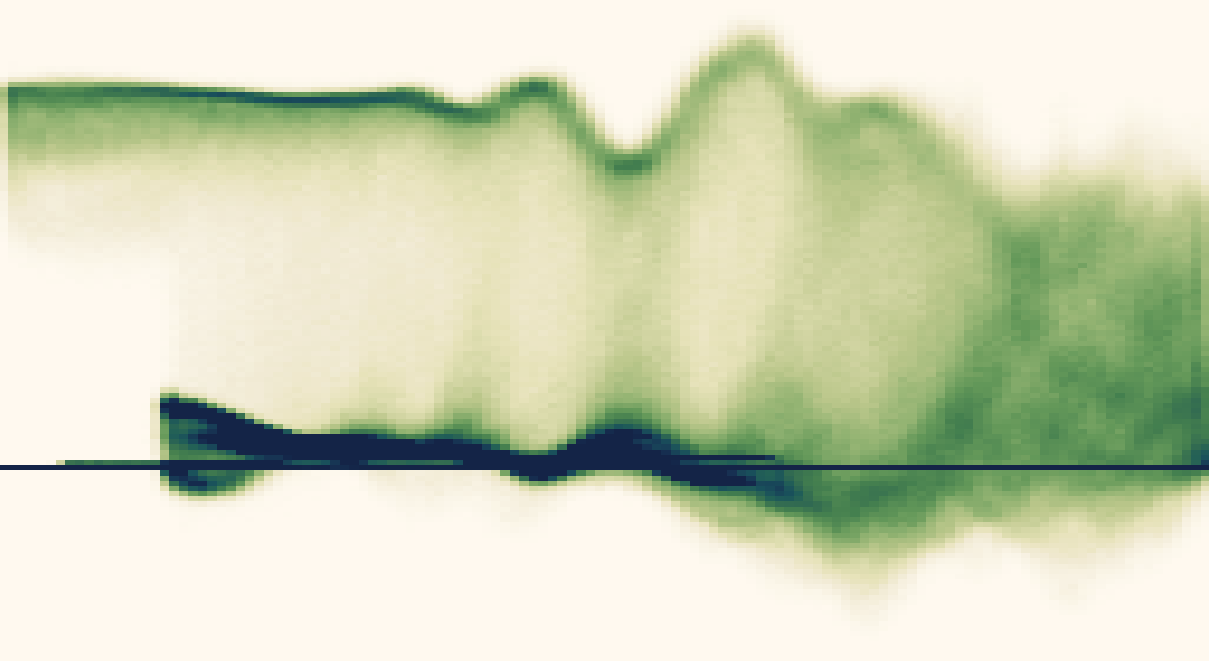}
	}
	\hfill
	\subfloat[PCP from samples]{%
		\includegraphics[height=22.5mm]{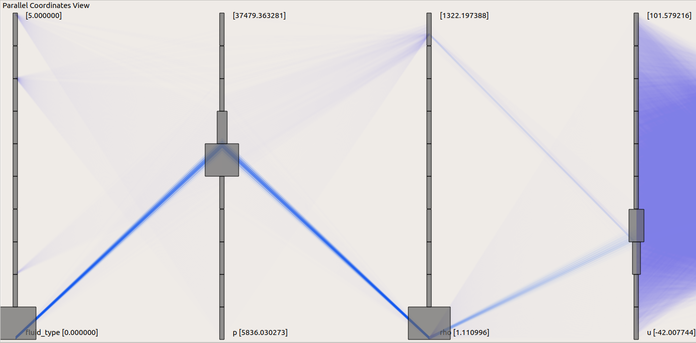}
	}
	\hfill
	\subfloat[Histogram from all particles]{%
		\includegraphics[height=22.5mm]{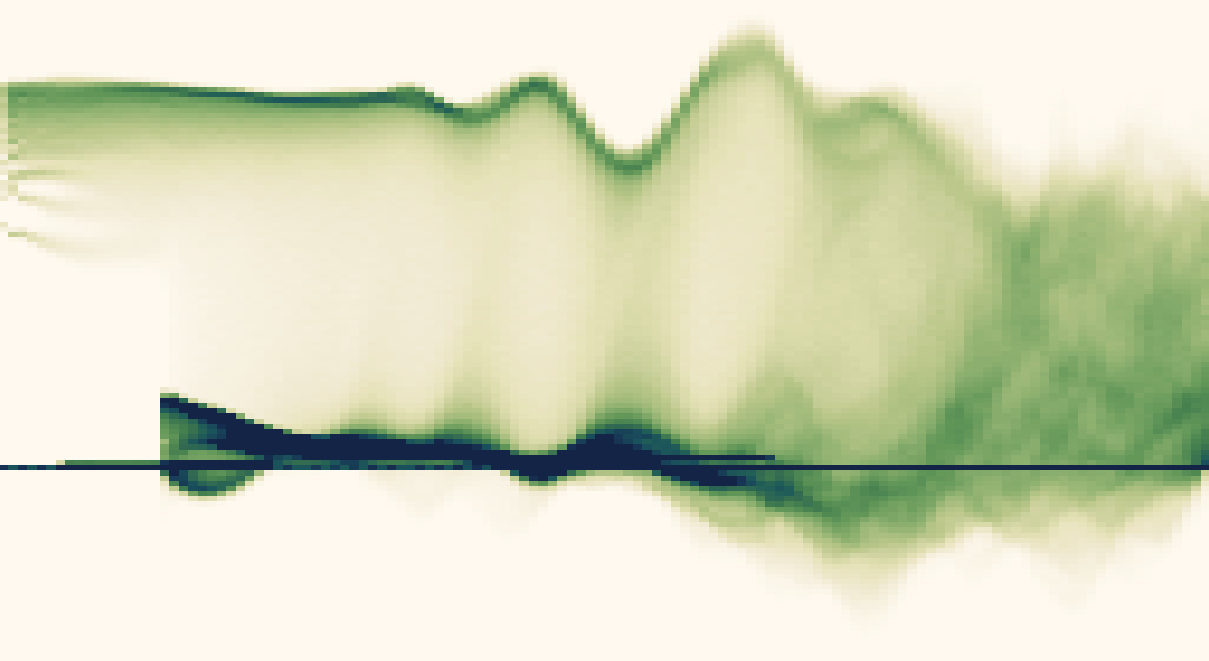}
	}
	\hfill
	\subfloat[PCP from all particles]{%
		\includegraphics[height=22.5mm]{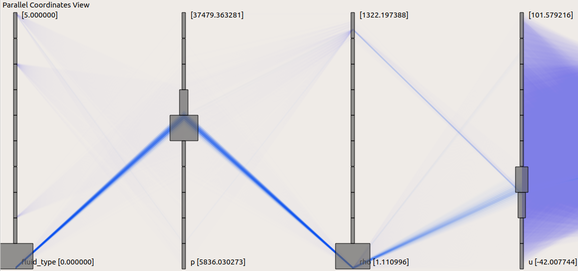}
	}
	\vspace{-1mm}
	\caption{Visualization of a spray nozzle using our approach with the $k$-means 32.000 clustering by splatting 3D Gaussians (a) and by drawing samples from the GMMs to create a 2D histogram (c) and a parallel coordinate plot (d). In (b), (e), and (f) the corresponding visualizations using the original SPH particle data are shown.}
	\label{fig:SprayNozzleSamples}
	\vspace{-5mm}
\end{figure*}

The Illustris simulation~\cite{Nelson2015} is a large-scale cosmological hydrodynamical simulation of galaxy formation that aims to predict both dark and baryonic components of matter.
In detail, the dynamics of dark matter and gas are simulated with the quasi-Lagrangian code AREPO~\cite{Springel2009}, which employs an unstructured Voronoi tessellation of the domain. After simulation, only the center points of the Voronoi cells are kept and are referred to as particles.
Since the simulation has been run in different resolutions and we want to show both dark and baryonic matter, we discuss multiple separate datasets as shown in \autoref{tbl:IllustrisData}. We compare the data based on the 100th time step without descendant or ancestor information.

The Illustris datasets have been clustered into halos using a domain specific approach. The sizes of clusters are extremely irregular and range in between a single particle and up to millions of particles per cluster. Since we cannot fit a GMM to very small clusters, we fit a single Gaussian for clusters of size below $20$.

With our approach, we are able to interactively visualize and explore these massive datasets \revi{that might not even fit into memory otherwise}. \autoref{fig:teaser} shows several interactive, linked views of the Illustris-1 dataset. We have brushed the x-, y-, and z-axis in the parallel coordinate plot (d). The brushed regions (green) are then highlighted in red in all other views. The density-based views are free of clutter and clearly show trends and correlations between the dimensions. For example, the parallel coordinate plot in (d) indicates that the brushed values have velocity components that are distributed around zero and are linearly correlated.
The spatial visualization depicts 5.3 million clusters that we render and navigate interactively.

\autoref{fig:Illustris3_Comparison} compares our probabilistic summaries with the original particle data of Illustris-3 Gas. \revi{Note that the interactive visualization of Illustris-1 and 2 is not possible on our system due to their data sizes.}
Although we clearly miss some details in the spatial visualization, we still manage to convey the general structure of the data and the distribution of color-mapped values. Whilst sorting and rendering all $16$ million particles as isotropic Gaussians takes \SI{61}{\milli\second} on our system, the clusters require only \SI{2}{\milli\second}. Note that for this dataset, the $110,000$ clusters are represented by a total of $357,512$ Gaussians in 3D\@.
In \autoref{fig:Illustris3_Comparison}(c), we have brushed a spatial region on the right side which is consequently put into focus. In (d) we have loaded the original particle data of the brushed clusters. Note that all of the linked views are also updated by this operation. Since we only load an additional $240,000$ particles, the interactive visualization still takes only \SI{3.8}{\milli\second} to render. Although other forms of level of detail are possible, this is a powerful way to drill-down from an overview to a detailed view of the data.

Since the clusters split and merge over time in this dataset, we have precomputed the transfer coefficients for time-dependent visualizations. \autoref{fig:Illustris3_TimeHist_Details} depicts a time-histogram of a selected attribute over several time steps. We have brushed the 94th time step, which is reflected in all views in the current, 100th time step. The spatial visualization in (b) highlights those brushed values. Due to our brush, mostly smaller clusters are in focus.
This brushing and linking over time thus allows exploring the selected dimension and its time-dependent behavior.
With our data summaries, we can compute the whole time-histogram in just under a second. In comparison, computing a time-histogram from the particle data takes \SI{34}{\second} and is severely I/O limited since the complexity scales with respect to the number of particles instead of the amount of clusters.

\subsection{Spray Nozzle}
\label{sec:Results:SprayNozzle}

\begin{table}
	\centering
	\caption{Overview of the spray nozzle dataset. We show the absolute summary size, relative to the original data size, the average number of GMM components, and the average Wasserstein distance.}
	\vspace{-1mm}
	\label{tbl:SprayNozzleData}
	\resizebox{0.95\linewidth}{!}{
		\begin{tabular}{ccccc}
			\toprule
			\# Clusters & Summaries & Rel. size & GMM comp. & Wasserstein dist. \\
			\midrule
			2,000 & 6.7 MB & \num{0.006}\% & $1.7 \pm 1.19$ & \num{5.54e-5} \\
			8,000 & 19.9 MB & \num{0.017}\% & $1.4 \pm 0.85$ & \num{1.57e-5} \\
			32,000 & 47.5 MB & \num{0.036}\% & $1.2 \pm 0.61$ & \num{4.64e-6} 	\\
			\bottomrule
	\end{tabular}
}
	\vspace{-4mm}
\end{table}

We have applied our technique to a smoothed particle hydrodynamics (SPH) dataset of a fuel spray nozzle simulation~\cite{Chaussonnet2018}. In the context of renewable energy production, biomass is converted into fuel by a gasification process. The quality of the spray is analyzed since it is critical for the efficiency of the gasification. However, the size of the time-dependent data prevents the usage of common interactive visual analysis techniques.
In detail, the dataset contains about 43 million particles per time step. Each particle contains a position, velocity, pressure, density, and fluid type for a total of 9 separate dimensions. The fluid type describes four different categories, including fluid, gas, and two types of boundaries.

We have partitioned the data using a $k$-means clustering based on the spatial position, fluid type, and velocity magnitude. \autoref{tbl:SprayNozzleData} shows the data size reduction and average number of GMM components for different numbers of clusters. For this dataset, we fix the maximum number of GMM components to $6$.
The size of the data summaries increases with the number of clusters. At the same time, the average number of GMM components decreases. This shows that the number of GMM components adapts to the less complex clusters. Moreover, the average Wasserstein distance is reduced for a larger number of clusters.

\autoref{fig:SprayNozzleSamples} depicts several visualizations created from our representation and from the original SPH data. The spatial visualizations in (a) and  (b) depict velocity in $u$-direction. Our approach does lose some details, especially on the finer structures on the right side of the cylindrical domain.
\revi{Since item-based visualizations of 43 million particles suffer from strong overdraw and visual clutter, density-based visualizations are preferable for this dataset. These are fast and efficient to create using our representation that is already an estimate of density.}
The 2D histogram in (b) and the parallel coordinate plot in (c) have been created from samples drawn from the GMMs. Compared to the reference plots in (e) and (f), we achieve nearly identical results. Moreover, it is possible to vary the number of samples, which could be used to create less cluttered visualizations, e.g.\ for scatter and parallel coordinate plots.

We represent the fluid type, i.e.\ the categorical dimension, by interpreting it as a scalar dimension. This is possible since the data only consists of four fluid types that we model using an appropriate number of Gaussian components. We could have increased the maximum number of components for all marginal distributions containing a categorical dimension, but this was not necessary for this dataset.
Although a small number of categories is common in multiphase fluid simulations, in general, representing categorical dimensions with GMMs does not scale.

\subsection{Hurricane Isabel}
\label{sec:Results:Isabel}

\begin{figure*}
	\centering
	\subfloat[Original data]{%
		\includegraphics[width=0.3\linewidth]{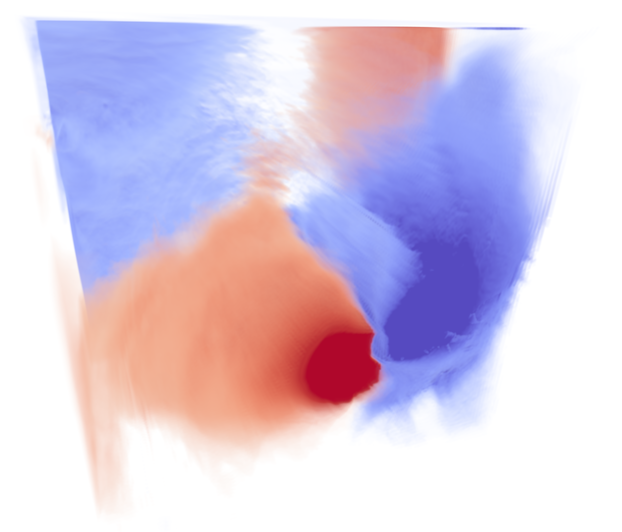}
	}
	\hfill
	\subfloat[Our low-dim. GMMs]{%
		\includegraphics[width=0.3\linewidth]{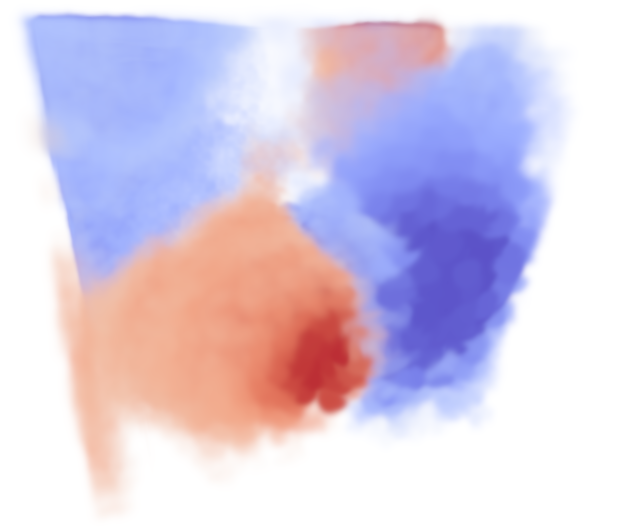}
	}
	\hfill
	\subfloat[High-dim. GMMs]{%
		\includegraphics[width=0.3\linewidth]{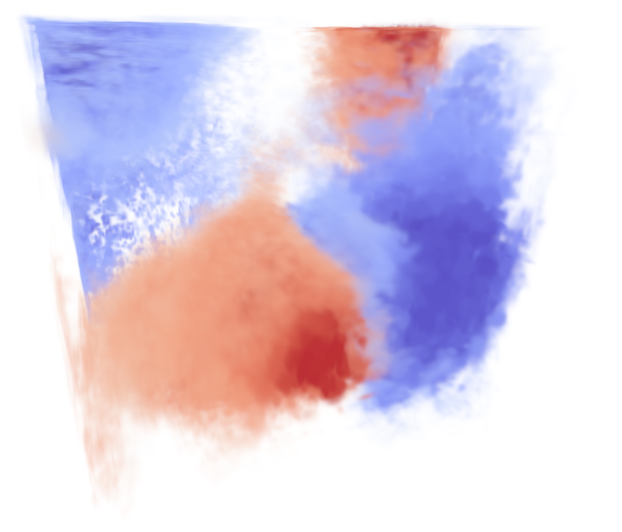}
	}
	\hfill
	\includegraphics[width=6.4mm]{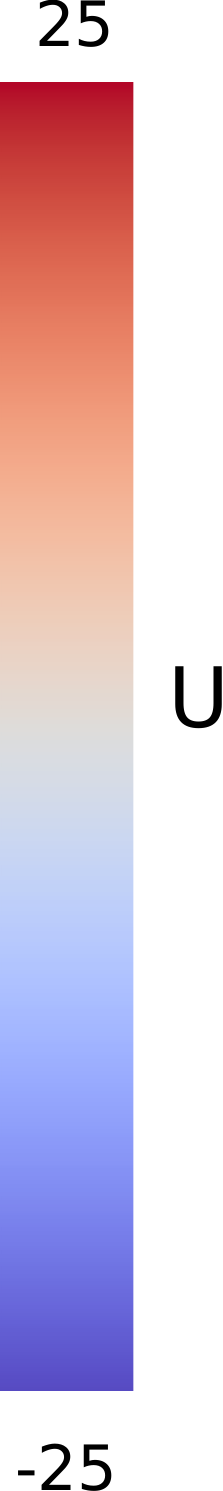}
	\vspace{-1mm}
	\caption{Visualization of wind speed from west to east (U) in the Hurricane Isabel data by splatting the original data (a), with the $k$-means 16.000 clustering of the low-dimensional model (b), and the high-dimensional model (c). }
	\label{fig:Isabel_Vis}
	\vspace{-3mm}
\end{figure*}

\begin{table}
	\centering
	\caption{Overview of the different clustering procedures of the Hurricane Isabel dataset. We show the resulting absolute and relative data size and the average Wasserstein distance. }
	\vspace{-1mm}
	\label{tbl:IsabelData}
	\resizebox{\linewidth}{!}{
		\begin{tabular}{llcccc}
			\toprule
			Model & Clustering & \# Clusters & Summaries & Rel. size & Wasserstein dist.\\
			\midrule
			Our & Blocks & 1,000 & 12.1 MB & 1.4\% & \num{1.20e-4} \\
			HD & Blocks & 1,000 & 5.2 MB & 0.6\% & \num{1.21e-4} \\
			\midrule
			Our & Blocks & 8,000 & 82.6 MB & 9.0\% & \num{1.58e-5} \\
			HD & Blocks & 8,000 & 34.7 MB & 4.8\% & \num{1.56e-5} \\
			\midrule
			Our & Blocks & 16,000 & 146.8 MB & 14.8\% & \num{8.49e-6} \\
			HD & Blocks & 16,000 & 50.0 MB & 5.1\% & \num{8.24e-6} \\
			\midrule
			Our & $k$-means & 1,000 & 12.1 MB & 1.4\% & \num{1.23e-4} \\
			HD & $k$-means & 1,000 & 5.4 MB & 0.6\% & \num{1.26e-4} \\
			\midrule
			Our & $k$-means & 8,000 & 80.5 MB & 8.8\% & \num{1.66e-5} \\
			HD & $k$-means & 8,000 & 33.8 MB & 3.7\% & \num{1.57e-5} \\
			\midrule
			Our & $k$-means & 16,000 & 143.9 MB & 14.7\% & \num{8.75e-6} \\
			HD & $k$-means & 16,000 & 48.5 MB & 5.0\% & \num{8.01e-6} \\
			\bottomrule
	\end{tabular}}
	\vspace{-4mm}
\end{table}

The Hurricane Isabel dataset is an atmospheric simulation from the IEEE Visualization Contest 2004, produced by the Weather Research and Forecast (WRF) model. Besides an implicit spatial position and a velocity vector, the time-dependent dataset contains 9 additional scalar quantities on a uniform grid of size $500 \times 500 \times 100$.
Since we formulated our approach for scattered data and did not consider the important but special case of gridded data, we disregard the topology of the dataset. 

To apply our approach, we either define clusters through uniform blocks or apply $k$-means clustering based on the spatial position and velocity magnitude. Both clustering procedures require a fixed number of clusters as input.
Independent from the clustering, we always store 3D distributions for the spatial position and the velocity vector and compute the respective 2D marginal distribution from these. Apart from that, we model and store all pairwise 2D distributions and all $15$ one-dimensional distributions. Additionally, we compare our representation to modeling each cluster with a high-dimensional Gaussian mixture model.

\autoref{tbl:IsabelData} shows the data summaries we have created with both clustering procedures, with different cluster sizes, with our approach and using high-dimensional Gaussian mixture models. We have chosen a maximum number of $6$ GMM components for the low-dimensional and $32$ for the high-dimensional models to achieve a comparable quality. Note that creating the high-dimensional model took nearly \num{43} hours, cf.~\autoref{tbl:Preprocessing}.
Both approaches can model the data well even though some dimensions are quite challenging.
The high-dimensional model performs surprisingly well for this dataset, considering the dimensionality, which is due to high correlations in the dimensions.
The low-dimensional representation requires more storage since it cannot make use of these higher-dimensional correlations. In both cases, the two clustering procedures lead to similar results. 

\autoref{fig:Isabel_Vis} shows a visualization of wind speed from west to east, i.e.\ $u$-velocity, by splatting the original data and with our approach. Although our representation loses some details, it conveys the major features of the dataset. Whilst the low-dimensional representation models the spatial position separately, the high-dimensional GMM takes \revi{correlations between} all dimensions into account. The marginal distribution of the spatial positions is thus also influenced by the other dimensions, which leads to the artifacts in \autoref{fig:Isabel_Vis}(c).
\revi{This reduces trust in the high-dimensional model since it is unclear if these correlations actually exist in the data or not.}
Lastly, the high-dimensional model contains over five times the amount of Gaussian components, which increases the complexity of all visualizations. 
\revi{In comparison, our representation consists of low-dimensional models that are easier to understand and more robust.}

\subsection{Performance}
\label{sec:Results:Perf}

\begin{table}
	\centering
	\caption{Performance of visualizations with our data summaries.}
	\vspace{-1mm}
	\label{tbl:Performance}
	\resizebox{\linewidth}{!}{
		\begin{tabular}{lccccc}
			\toprule
			Dataset & \multicolumn{2}{c}{Splatting} & PCP & Density & Brushing \\
			\cmidrule{2-3}
			& Sorting & OIT & & $(x, u)$\\
			\midrule
			Illustris-3 Gas & 3.9ms & 4.3ms & 439ms & 1.0ms & 4ms \\
			Illustris-2 DM & 31ms & 14ms & 421ms & 3.6ms & 28ms \\
			Illustris-1 DM & 196ms & 28ms & 1241ms & 11.2ms & 160ms \\
			Hurricane Isabel \small{k=8000} & 4.6ms & 20ms & 99ms & 1.1ms & 2ms \\
			Spray Nozzle \small{k=8000} & 4.4ms & 23ms & 47ms & 1.4ms & 2ms \\
			\bottomrule
	\end{tabular}}
	\vspace*{-4mm}
\end{table}

Our evaluations were performed on an Intel i7-6700 with 32~GB of system memory and an NVIDIA GTX 1080 Ti graphics card providing 11~GB of video memory. For GPU acceleration, we make use of both CUDA for general purpose computations and OpenGL for rendering. For our spatial visualization, we have used a screen resolution of $1920\times1080$. The resolution of our 2D density plots was $200\times 200$ and $800 \times 300$ for the parallel coordinate plot (PCP).

Timings for several visualizations are shown in \autoref{tbl:Performance}. In general, our prototype allows interactive navigation and creation of all visualizations introduced above.
The Illustris 1 and 2 datasets are the most demanding, due to the large number of clusters. Note that the performance of our approach scales with the number of clusters and Gaussian components, not the original data size.
%
The order-independent transparency (OIT) approach performs very well on the cosmological datasets compared to the back-to-front splatting using sorting. Note that the speed varies depending on the number of covered pixels. The sorting approach is faster on the smaller and spatially more compact datasets. 

We create our probabilistic data summaries in a preprocessing step using the Python scikit-learn library. This process is trivial to parallelize since all time steps, clusters, and distributions can be processed independently.
Due to inherent restrictions imposed by our Python prototype, an implementation in a native language is expected to be significantly faster.
The measurements for our prototype are shown in \autoref{tbl:Preprocessing}.
Our fast GMM component estimation (\autoref{sec:Model:GMMs}) leads to a significant speedup. In the supplementary material, we show that a slight error is introduced by this approximation.
Lastly, computing high-dimensional GMMs requires significantly more preprocessing time, making it unsuited for use in practice. Note that our approximations for a fast estimation of GMM components cannot be used for the high-dimensional data.

\begin{table}
	\centering
	\caption{Measurements of the data summary preprocessing.}
	\vspace{-1mm}
	\label{tbl:Preprocessing}
	\resizebox{\linewidth}{!}{
		\begin{tabular}{lccc}
			\toprule
			Dataset & Our GMMs & Low-dim. GMMs & High-dim. GMMs\\
			\midrule
			Hurricane Isabel \small{k=1000} & 2h 54m & 9h 31m & 42h 55m \\
			Spray Nozzle \small{k=2000} & 1h 43m & 8h 13m & 14h 51m \\
			\bottomrule
	\end{tabular}}
	\vspace*{-4mm}
\end{table}

\section{Conclusion}

In this paper, we introduce probabilistic data summaries for multivariate scattered data. They enable the interactive visual analysis of large datasets that would not be possible otherwise \revi{due to limitations of memory or compute}. Although our data representation is a simplified model of the data, we inform the user about this uncertainty and present a level-of-detail and outlier visualization for more detailed investigations.

The core insight of our approach is that we only have to model combinations of low-dimensional distributions for visual analysis, which avoids the complexity of modeling high-dimensional distributions.
Although the data must be clustered, we do not make any restrictive assumptions about the clustering procedure. In fact, our evaluation shows that the impact of the clustering on the quality of the representation is less pronounced than expected and is largely offset by the adaptive modeling of GMMs.

In the future, we want to improve the scalability of our approach even further by adding a level-of-detail approach based on a hierarchical clustering of the data. By interactively selecting the appropriate detail, it should be possible to interactively explore massive datasets even on mobile devices and seamlessly scale up to powerful workstations.

\acknowledgments{
The Spray Nozzle dataset is due to the Institute of Thermal Turbomachinery (ITS) at the Karlsruhe Institute of Technology (KIT).
The Hurricane Isabel data is courtesy of NCAR\@, and the U.S.\@ National Science Foundation (NSF).}

\bibliographystyle{abbrv-doi}

\bibliography{bibliography2}
\end{document}


\maketitle

	\section{3D Gaussian Ray Integration}
	
	As motivated in the paper, we integrate a trivariate Gaussian distribution along a ray $\RayOrigin + \RayVariable \RayDirection$ starting at $\RayOrigin \in \R^3$ in normalized direction $\RayDirection \in \R^3$ with $\RayVariable \in \R$.
	The Gaussian is given by its mean $\SingleGaussianMean\in\R^3$ and covariance $\SingleGaussianCovariance\in\R^{3\times3}$. To derive a general solution, we integrate over $[\IntervalBegin, \IntervalEnd]$ by substituting the ray equation into the trivariate Gaussian distribution:
	\begin{equation}
	\GaussianIntegral{\IntervalBegin}{\IntervalEnd} := \int_{\IntervalBegin}^{\IntervalEnd} \frac{1}{\sqrt{|2\pi\SingleGaussianCovariance|}} \exp \left(-\frac{\Transpose{(\RayOrigin + \RayVariable\RayDirection - \SingleGaussianMean)} \SingleGaussianCovariance^{-1} (\RayOrigin + \RayVariable\RayDirection - \SingleGaussianMean)}{2} \right)\,\textrm{d}\RayVariable.
	\end{equation}
	Note that $|2\pi\SingleGaussianCovariance| = (2\pi)^3 |\SingleGaussianCovariance|$ for trivariate Gaussians, which we prefer due to its compactness. We start by simplifying the equation:
	\begin{equation}
	\begin{split}
	\GaussianIntegral{\IntervalBegin}{\IntervalEnd} & = \int_{\IntervalBegin}^{\IntervalEnd} \frac{1}{\sqrt{|2\pi\SingleGaussianCovariance|}} \exp \left(-\frac{\Transpose{((\RayOrigin - \SingleGaussianMean) + \RayVariable\RayDirection)} \SingleGaussianCovariance^{-1} ((\RayOrigin - \SingleGaussianMean) + \RayVariable\RayDirection)}{2} \right)\,\textrm{d}\RayVariable \\
	& =
	\int_{\IntervalBegin}^{\IntervalEnd}
	\frac{1}{\sqrt{|2\pi\SingleGaussianCovariance|}}
	\exp \left(-\Transpose{(\RayOrigin - \SingleGaussianMean)} \SingleGaussianCovariance^{-1} (\RayOrigin - \SingleGaussianMean) \right)
	\exp \left(- 2 \RayVariable \underbrace{\frac{1}{2}\Transpose{(\RayOrigin - \SingleGaussianMean)} \SingleGaussianCovariance^{-1} \RayDirection}_{\IntegralAuxiliaryOffsetDirection} \right)
	\exp \left(- \RayVariable^2 \underbrace{\frac{1}{2}\Transpose{\RayDirection} \SingleGaussianCovariance^{-1} \RayDirection}_{\IntegralAuxiliaryDirection} \right)\,\textrm{d}\RayVariable \\
	& =
	\frac{1}{\sqrt{|2\pi\SingleGaussianCovariance|}}
	\exp \left(-\Transpose{(\RayOrigin - \SingleGaussianMean)} \SingleGaussianCovariance^{-1} (\RayOrigin - \SingleGaussianMean) \right)
	\int_{\IntervalBegin}^{\IntervalEnd}
	\exp \left( -2 \RayVariable \IntegralAuxiliaryOffsetDirection - \RayVariable^2 \IntegralAuxiliaryDirection \right)\,\textrm{d}\RayVariable .
	\end{split}
	\end{equation}
	We substitute $\FirstSubstitutionVariable := \RayVariable + \frac{\IntegralAuxiliaryOffsetDirection}{\IntegralAuxiliaryDirection}$ and thus rewrite and simplify the integrand as follows:
	\begin{equation}
	\begin{split}
	\GaussianIntegral{\IntervalBegin}{\IntervalEnd} & =
	\frac{1}{\sqrt{|2\pi\SingleGaussianCovariance|}}
	\exp \left(-\Transpose{(\RayOrigin - \SingleGaussianMean)} \SingleGaussianCovariance^{-1} (\RayOrigin - \SingleGaussianMean) \right)
	\int_{\IntervalBeginFirstSubstitution}^{\IntervalEndFirstSubstitution}
	\exp \left(- 2 \left(\FirstSubstitutionVariable - \frac{\IntegralAuxiliaryOffsetDirection}{\IntegralAuxiliaryDirection} \right) \IntegralAuxiliaryOffsetDirection - \left( \FirstSubstitutionVariable - \frac{\IntegralAuxiliaryOffsetDirection}{\IntegralAuxiliaryDirection} \right) ^2 \IntegralAuxiliaryDirection \right)\,\textrm{d}\FirstSubstitutionVariable \\
	& =
	\frac{1}{\sqrt{|2\pi\SingleGaussianCovariance|}}
	\exp \left(-\Transpose{(\RayOrigin - \SingleGaussianMean)} \SingleGaussianCovariance^{-1} (\RayOrigin - \SingleGaussianMean) \right)
	\int_{\IntervalBeginFirstSubstitution}^{\IntervalEndFirstSubstitution}
	\exp \left(- 2 \FirstSubstitutionVariable \IntegralAuxiliaryOffsetDirection + 2 \frac{\IntegralAuxiliaryOffsetDirection^2}{\IntegralAuxiliaryDirection} - \FirstSubstitutionVariable^2 \IntegralAuxiliaryDirection + 2 \FirstSubstitutionVariable \IntegralAuxiliaryOffsetDirection - \frac{\IntegralAuxiliaryOffsetDirection^2}{\IntegralAuxiliaryDirection} \right)\,\textrm{d}\FirstSubstitutionVariable \\
	& =
	\underbrace{\frac{1}{\sqrt{|2\pi\SingleGaussianCovariance|}}
	\exp \left(-\Transpose{(\RayOrigin - \SingleGaussianMean)} \SingleGaussianCovariance^{-1} (\RayOrigin - \SingleGaussianMean) \right)
	\exp \left(\frac{\IntegralAuxiliaryOffsetDirection^2}{\IntegralAuxiliaryDirection} \right)}_{\IntegralAuxiliary}
	\int_{\IntervalBeginFirstSubstitution}^{\IntervalEndFirstSubstitution}
	\exp \left( -\FirstSubstitutionVariable^2 \IntegralAuxiliaryDirection \right) \,\textrm{d}\FirstSubstitutionVariable .\\
	\end{split}
	\end{equation}
	
	Now, we perform another substitution using $\SecondSubstitutionVariable := \FirstSubstitutionVariable \sqrt{\IntegralAuxiliaryDirection}$:
	\begin{equation}
	\GaussianIntegral{\IntervalBegin}{\IntervalEnd} = 
	\IntegralAuxiliary
	\frac{1}{\sqrt{\IntegralAuxiliaryDirection}}
	\int_{\IntervalBeginSecondSubstitution}^{\IntervalEndSecondSubstitution}
	\exp(-p^2) \,\textrm{d}\SecondSubstitutionVariable .
	\end{equation}
	The integrand can now be expressed by its antiderivative, which leads us to the following solution:
	\begin{equation}
	\GaussianIntegral{\IntervalBegin}{\IntervalEnd} = \IntegralAuxiliary
	\frac{1}{\sqrt{\IntegralAuxiliaryDirection}}
	\left[\frac{\sqrt{\pi}}{2} \erf(p) \right]_{\IntervalBeginSecondSubstitution}^{\IntervalEndSecondSubstitution}.
	\end{equation}
	Although no closed-form solution exists for the error function, fast and accurate numerical approximations of this well known function are available. Lastly, when integrating over $(-\infty, \infty)$, the error function disappears:
	$$ \left[ \frac{\sqrt{\pi}}{2} \erf(p) \right]_{-\infty}^{\infty} = \sqrt{\pi},$$
	which leads to
	\begin{equation}
		\GaussianIntegral{-\infty}{\infty} = \IntegralAuxiliary \frac{\sqrt{\pi}}{\sqrt{\IntegralAuxiliaryDirection}}.
	\end{equation}
	
	
\section{Fast Selection of GMM Components}

In the main paper, we propose a fast selection of GMM components by formulating lower und upper bounds and using subsampling. In \autoref{fig:SprayNozzleGMMComparison}, we illustrate the impact of this approximation on the Spray Nozzle (a) and the Hurricane Isabel (b) dataset compared to a brute-force selection of GMM components. 
Using only the bounds does not introduce a measurable error. However, the subsampling leads to a slightly increased error.
Note that for subsampling we always take a fixed amount of \si{200} samples from each cluster.
By increasing this fixed number of samples, we lower this error, at the cost of additional computational effort.

\begin{figure}[p]
	\centering
	\subfloat[Spray Nozzle (k=2000)]{%
		\includegraphics[width=0.49\linewidth]{ApproximComparison/150errors.pdf}
	}
	\hfill
	\subfloat[Hurricane Isabel (k=1000)]{%
		\includegraphics[width=0.49\linewidth]{ApproximComparison/40errors.pdf}
	}
	\caption{Comparison of our fast selection of GMM components with the brute-force approach on the Spray Nozzle (a) and the Hurricane Isabel (b) dataset.}
	\label{fig:SprayNozzleGMMComparison}
\end{figure}
	
\section{Additional Results}

In this section, we show additional results for the datasets presented in the paper.

\subsection{Synthetic Dataset}
	
%

	\begin{figure}[p]
	\centering
	\subfloat[Error of 1D dimensions]{%
		\includegraphics[height=44mm]{SyntheticError/Synthetic_clustering_comparison.pdf}
	}
	\subfloat[Probability density function]{%
		\includegraphics[height=44mm]{SyntheticError/exp_highdim_pdf.pdf}
	}
	\subfloat[Cumulative distribution function]{%
		\includegraphics[height=44mm]{SyntheticError/exp_highdim_cdf.pdf}
	}
	\caption{For the synthetic dataset, we compare our data summaries to high-dimensional GMMs (a). In (b), the exponentially distributed dimension from the high-dimensional GMMs is shown. The corresponding CDF in (c) indicates a high error.}
	\label{fig:HighDimError}
	\end{figure}

\autoref{fig:HighDimError} compares our data summaries to high-dimensional Gaussian mixture models. In (a), the mean Wasserstein distance of all 1D dimensions is shown. Although the error is comparable, the high-dimensional GMMs lead to a higher error for the more complex dimensions, such as the exponentially distributed dimension shown in (b) and (c).

\subsection{Spray Nozzle}

In \autoref{fig:SprayNozzleComparison}, we compare our data summaries to those of high-dimensional GMMs. In (a), the error of our and the high-dimensional data model is shown. The error is comparable between both approaches and low in absolute size. The table in (b) shows that the high-dimensional data model is smaller in size. However, the high-dimensional model requires extremely expensive preprocessing and the large amount of used GMM components affects the performance of all visualizations.


\begin{figure}[p]
	\centering
	\subfloat[Error of 1D dimensions]{%
		\raisebox{0pt}{\includegraphics[height=50mm]{SprayNozzleComparison/SprayNozzle_clustering_comparison.pdf}}
	}%
	\hfill
	\subfloat[Overview of the data models]{%
	\raisebox{50pt}{\resizebox{0.52\linewidth}{!}{%
		\begin{tabular}{lcccc}
			\toprule
			Model & \# Clusters & Size & GMM comp. & Wasserstein dist. \\
			\midrule
			Our & 2,000 & 6.7 MB & $1.7 \pm 1.19$ & \num{5.54e-5} \\
			HD & 2,000 & 3.7 MB & $21.6 \pm 3.8$ & \num{5.50e-5} \\
			\midrule
			Our & 8,000 & 19.9 MB & $1.4 \pm 0.85$ & \num{1.57e-5} \\
			HD & 8,000 & 9.2 MB & $13.5 \pm 3.5$ & \num{1.59e-5} \\
			\midrule
			Our & 32,000 & 47.5 MB  & $1.2 \pm 0.61$ & \num{4.64e-6} \\
			HD & 32,000 & 13.6 MB & $4.9 \pm 2.86$ & \num{6.24e-6} \\
			\bottomrule
	\end{tabular}}}
	}%
	\caption{We compare our data summaries to modeling high-dimensional GMMs.}
	\label{fig:SprayNozzleComparison}
\end{figure}